\documentclass[prodmode]{acmsmall-mod}

\makeatletter
\def\@acmYear{2012}
\def\@acmMonth{11}
\def\@acmVolume{V}
\def\@acmNumber{N}
\def\@acmArticle{A}

\let\subcaption\@undefined	
\makeatother

\usepackage{caption}
\usepackage{subcaption}
\usepackage{paralist}

\usepackage{url}
\usepackage{graphicx}

\usepackage{amssymb}
\usepackage{array}
\usepackage{enumerate}
\usepackage{booktabs}
\usepackage{verbatim}
\usepackage{fancyvrb,relsize}

\usepackage{xcolor}
\definecolor{labelcolor}{cmyk}{0.22,0.10,0.10,0.10}
\definecolor{listbackgroundcolor}{cmyk}{0.10,0.10,0.05,0.05}

\usepackage{tikz}
\usetikzlibrary{arrows,shapes,snakes,automata,backgrounds,petri}
\usetikzlibrary{calc}
\usetikzlibrary{fit}
\usetikzlibrary{matrix}
\usetikzlibrary{positioning}
\usetikzlibrary{decorations}
\usetikzlibrary{decorations.markings}

\usepackage{times}
\usepackage{amsmath, amssymb}
\usepackage{latexsym}
\usepackage{graphicx}
\usepackage{epstopdf}
\DeclareMathAlphabet{\mathsl}{OT1}{ptm}{m}{sl}

\newcommand{\ifsubmit}[1]{}

\usepackage{xspace}
\newcommand{\etal}{{et al.\@\xspace}}

\newcommand{\be}{\begin{itemize}}
\newcommand{\ee}{\end{itemize}}
\newcommand{\bn}{\begin{enumerate}}
\newcommand{\en}{\end{enumerate}}
\newcommand{\bc}{\begin{center}}
\newcommand{\ec}{\end{center}}
\newcommand{\bl}{\begin{flushleft}}
\newcommand{\el}{\end{flushleft}}
\newcommand{\bq}{\begin{quote}}
\newcommand{\eq}{\end{quote}}
\newcommand{\beq}{\begin{equation}}
\newcommand{\eeq}{\end{equation}}

\newcommand{\bmp}{\begin{minipage}}
\newcommand{\emp}{\end{minipage}}

\newcommand{\mypara}[1]{\medskip\noindent\fbf{#1}}

\newcommand{\bdeff}{\begin{definition}\rm} 
\newcommand{\edeff}{\end{definition}} 

\newtheorem{query}{Query}
\newcommand{\bqry}{\begin{query}\rm} 
\newcommand{\eqry}{\end{query}}

\newcommand{\bexa}{\begin{example}\rm} 
\newcommand{\eexa}{\end{example}}
\newcommand{\eeeexa}{\rule[-0.1mm]{1.0mm}{3mm}\end{itemize}\end{example}}
\newcommand{\eeeeexa}{\rule[-0.1mm]{1.0mm}{3mm}\end{itemize}\end{itemize}\end{example}}

\newcommand{\eenexa}{\rule[-0.1mm]{1.0mm}{3mm}\end{enumerate}\end{example}}
\newcommand{\eeee}{\end{itemize}\end{itemize}}

\newtheorem{axiom}{Axiom}
\newcommand{\baxi}{\begin{axiom}\rm\flushleft}
\newcommand{\eaxi}{\end{axiom}}

\newcounter{characteristics}
\newcommand{\bch}{\begin{exfacts}{C}\setcounter{enumi}{\value{characteristics}}\renewcommand{\theenumi}{$\mathbf{C\emph{-}\arabic{enumi}}$}}
\newcommand{\ech}{\setcounter{characteristics}{\value{enumi}}\renewcommand{\theenumi}{
\arabic{enumi}.}\end{exfacts}}

\newcounter{observations} 
\newcommand{\bobs}{\begin{exfacts}{O}\setcounter{enumi}{\value{observations}}\renewcommand{\theenumi}{$\mathbf{Observation_\arabic{enumi}}$}}
\newcommand{\eobs}{\setcounter{observations}{\value{enumi}}\renewcommand{\theenumi}{
\arabic{enumi}.}\end{exfacts}}

\newcounter{consequences} 
\newcommand{\bcons}{\begin{exfacts}{O}\setcounter{enumi}{\value{consequences}}\renewcommand{\theenumi}{$\mathbf{Consequence_\arabic{enumi}}$}}
\newcommand{\econs}{\setcounter{consequences}{\value{enumi}}\renewcommand{\theenumi}{
\arabic{enumi}.}\end{exfacts}}

\newcommand{\bcor}{\begin{corollary}\rm\flushleft}
\newcommand{\ecor}{\end{corollary}}

\newtheorem{postulate}{Postulate}
\newcommand{\bpos}{\begin{postulate}\rm\flushleft}
\newcommand{\epos}{\rule[-0.25mm]{1.75mm}{3mm}\end{postulate}}

  \newcommand{\speciallabelsize}{\normalsize\rm}
\newenvironment{exfacts}[1]{\begin{list}{{\speciallabelsize \theenumi.}}{\usecounter{enumi}
        \settowidth{\labelwidth}{{\speciallabelsize #199}}
        \setlength{\leftmargin}{\labelwidth}
        \addtolength{\leftmargin}{2.0\labelsep}}}{\end{list}}
\newcounter{facts}
\newcommand{\bfact}{\begin{exfacts}{F}\setcounter{enumi}{\value{facts}}\renewcommand{\theenumi}{F\arabic{enumi}}}
\newcommand{\efact}{\setcounter{facts}{\value{enumi}}\renewcommand{\theenumi}{
\arabic{enumi}.}\end{exfacts}}

\newcounter{rules}
\newcommand{\brule}{\begin{exfacts}{R}\setcounter{enumi}{\value{rules}}\renewcommand{\theenumi}{R\arabic{enumi}}}
\newcommand{\erule}{\setcounter{rules}{\value{enumi}}\renewcommand{\theenumi}{
\arabic{enumi}.}\end{exfacts}}

\usepackage[scaled=0.9]{helvet}

\newcommand{\fbf}{\textbf}

\newcommand{\fsf}[1]{\textsf{\small{#1}}}

\newcommand{\fsc}{\textsc}
\newcommand{\fsl}{\textsl}

\newcommand{\msf}{\mathsf}
\DeclareMathAlphabet{\mathsl}{OT1}{ptm}{m}{sl}

\newcommand{\C}{\fsf{C}}

\newcommand{\F}{$\checkmark$}
\newcommand{\Q}{$-$}
\newcommand{\V}{$\times$}

\newcommand{\bexan}[1]{\begin{example}[#1]\rm} 

\usepackage{wasysym}

\begin{document}
\markboth{Chopra and Singh}{Interaction-Oriented Software Engineering}
\title{Interaction-Oriented Software Engineering:  Concepts and Principles}

\author{%
  AMIT K.~CHOPRA \affil{Lancaster University}
  MUNINDAR P.~SINGH \affil{North Carolina State University}
}

\begin{abstract}

  Following established tradition, software engineering today is
  rooted in a conceptually centralized way of thinking.  The primary
  SE artifact is a specification of a machine---a computational
  artifact---that would meet the (elicited and) stated requirements.
  Therein lies a fundamental mismatch with (open) sociotechnical
  systems, which involve multiple autonomous social participants or
  \emph{principals} who interact with each other to further their
  individual goals.  No central machine governs the behaviors of the
  various principals.

  We introduce Interaction-Oriented Software Engineering (IOSE) as an
  approach expressly suited to the needs of open sociotechnical
  systems.  In IOSE, specifying a system amounts to specifying the
  interactions among the principals as \emph{protocols}.  IOSE
  reinterprets the classical software engineering principles of
  modularity, abstraction, separation of concerns, and encapsulation
  in a manner that accords with the realities of sociotechnical
  systems.  To highlight the novelty of IOSE, we show where well-known
  SE methodologies, especially those that explicitly aim to address
  either sociotechnical systems or the modeling of interactions among
  autonomous principals, fail to satisfy the IOSE principles.

\end{abstract}



%

\begin{bottomstuff}
Amit~K.~Chopra:
School of Computing and Communications, Lancaster University,
Lancaster LA1 4WA, United Kingdom.
\fsf{a.chopra1@lancaster.ac.uk}.
Munindar~P.~Singh:
Department of Computer Science, North Carolina State University,
Raleigh, NC 27695-8206, USA.
\fsf{m.singh@ieee.org}
\end{bottomstuff}

\maketitle

\section{Introduction}
\label{sec:introduction}
We define a \emph{sociotechnical system} as one involving interactions
between \emph{autonomous} social entities such as people and
organizations mediated by technical components.  By emphasizing the
autonomous nature of social entities, our definition generalizes over,
yet more precisely captures, the traditional connotation of the
interaction between humans and societal infrastructure.  Our
definition contrasts with one of the conventional uses of the term,
which covers any interaction between people and computers.

Researchers in software engineering (SE) picked up the theme of
sociotechnical systems in two major directions: (1) how to model a
sociotechnical system as a combination of social and software
components, as in
\cite{bryl:socio-technical-systems:2009,yu:i-star:1997,Yu+11}; and (2)
how to elicit, model, and manage the requirements of the social
components so that suitable software components may be designed
\cite{baxter:sts:2011,goguen:re:1994,mylopoulos:nonfunctional-requirements:1992}.
Often, there is substantial overlap between the two directions, for
example, as in i* \cite{yu:i-star:1997}.  Baxter and Sommerville
\shortcite{baxter:sts:2011} conceptualize a sociotechnical system as
one that \emph{actively pursues goals} in an organizational setting;
that is, a sociotechnical system is one actor: a single locus of
control.  Therefore, we refer to such systems as being
\emph{conceptually centralized}.  The centralized conception is shared
by current SE approaches.

For clarity, we reserve the term \emph{principal} to refer to an
autonomous social party who participates in a system at runtime and
the term \emph{stakeholder} to refer to one who originates some of the
requirements.  In general, any principal would also be a stakeholder
in the system (suitably abstracted via the notion of \emph{roles}).

We restrict our attention to systems whose membership and structure
can change dynamically.  Such \emph{open} sociotechnical systems are
properly viewed as \emph{societies} of principals: we cannot build a
``complete'' system but can only specify how its members may
\emph{interact}, leaving as out of our scope the engineering of the
members themselves.  Thus whether the members \emph{comply} with the
specified interactions is crucial.  Because of its emphasis on
interaction, we term our approach interaction-oriented software
engineering or \emph{IOSE}.  We claim that IOSE is necessary and show
that the centralized conception is not applicable in sociotechnical
systems.

\subsection{Example: Healthcare}
\label{sec:healthcare}

Common settings such as business services and social computing realize
open sociotechnical systems.  Consider Schield {\etal}'s
\shortcite{schield:healthcare:2001} description of a healthcare system
in the United States, including its key stakeholders and their
(presumed) objectives.

\begin{itemize}
\item \fsc{us society}: the collective public, private, and personal
  interests of the United States citizens.
\item \fsc{regulatory bodies}: public and private that make and
  enforce policies.
\item \fsc{mco}: managed care organizations (i.e., licensed insurance
  organizations), who offer and administer managed care plans.
\item \fsc{institutional providers}: hospitals and laboratories.
\item \fsc{clinical or professional providers}: individual doctors and
  communities of practice.
\item \fsc{employers}: fully insured and self-funded organizations who
  offer managed care plans to their employees.
\item \fsc{consumers}: the enrollees of managed care plans.
\item \fsc{medicaid or medicare beneficiaries}.
\end{itemize}

Schield {\etal} find that the stakeholders' objectives often conflict.
For example, payers must keep costs down whereas providers must
maximize revenue; insurers want to shift risk to providers and
consumers for matters of cost control whereas both providers and
consumers want protection from potentially catastrophic costs.
Consumers want shorter waiting periods for appointments with their
physicians whereas physicians want to increase their panel of
patients.  Further, Schield {\etal} point out how one principal can,
in pursuing its own objectives, compromise the objectives of others.
For example, making a patient wait longer may compromise the patient's
health, thereby increasing the cost of care and decreasing patient
satisfaction.  Regulatory bodies have the objective of making sure
MCOs and providers play by the rules and ensuring that cost, quality,
and access criteria are met but that conflicts with society's goals of
lower-cost healthcare.

Assume that the stakeholders for a healthcare system manage to resolve
their differences and specify a system that meets their stated
collective requirements.  Now, individual \emph{principals} such as
specific MCOs, hospitals, physicians, employees, and consumers can
join or leave the managed healthcare system of their own volition.
The principals act according to their own private business policies,
some of which may not have been envisaged by the stakeholders.  For
example, an MCO may have a private objective to acquire an independent
call center.

Further, principals would generally employ their own internal
information systems to support their interactions with other
principals.  For example, providers may set up appointment systems to
handle appointment-related communications with consumers.  Hospitals
and MCOs may use complex information systems that help process
payments from each other and the consumers.  MCOs may employ complex
actuarial and decision-support processes in handling claims.

Recognizing and addressing conflicts among stakeholders requirements
has been a longstanding area of research in software engineering.
Once the conflicts are addressed, through whichever means, engineers
would seek to model and realize a system that meets those
requirements.  This paper treats conflict analysis as out of scope and
instead focuses on identifying modeling assumptions and criteria that
capture sociotechnical systems more faithfully.

\subsection{Contributions and Organization}
This paper makes two main contributions.  First, it shows via
conceptual analysis why IOSE is (1) a novel paradigm and (2)
well-suited for the software engineering of open sociotechnical
systems.  Our argument proceeds as follows.  One, we show that the
foundational models of traditional SE are conceptually centralized.
This is because traditional SE concerns itself with the engineering of
what is conceptually a single machine
(Section~\ref{sec:machine-orientation}).  Two, we present the
conceptual model of a system in IOSE (Section~\ref{sec:IOSE}) and
discuss how it accommodates multiple perspectives.  Three, based on
the characteristics of sociotechnical systems, namely, autonomy,
accountability, and loose coupling, we formulate criteria that any SE
approach for them should meet and show that whereas
machine-orientation, and therefore traditional SE, fails the criteria,
IOSE meets them (Section~\ref{sec:eval-machine-IOSE}).

The second contribution of this paper lies in reinterpreting the broad
SE principles of modularity, encapsulation, abstraction, and the
separation of concerns in accordance with the above criteria
(Section~\ref{sec:principles-IOSE}).  The reinterpretation provides
the elements of a methodology for IOSE.  Further, we discuss in detail
well-known software methodologies that are motivated either by
sociotechnical systems or the modeling of interactions among
principals and show how they violate one or more of these principles
(Section~\ref{sec:prominent-approaches}).
Section~\ref{sec:literature} discusses additional literature that is
relevant to IOSE.  Section~\ref{sec:discussion} concludes the paper
with pointers to future directions.

\section{Traditional Software Engineering: Machines}
\label{sec:machine-orientation}
Figure~\ref{fig:RE-conceptual-model} (from Van Lamsweerde
\shortcite{vanLamsweerde:requirements:2009}) shows the traditional
conceptual model for SE.  The \emph{system as is} represents the
system with identified problems, inefficiencies, and limitations.  The
\emph{system to be}, whose \emph{objectives} are to avoid those
deficiencies, is to be engineered.  The idea is to understand the
problem domain in order to come up with a set of \emph{services},
\emph{constraints}, and \emph{assumptions} under which the stated
objectives would be met.  Some of these would be met by the
\emph{software to be} as part of the \fsl{system to be}; the rest
would be assigned as responsibilities to components in the
\emph{environment}, namely, \emph{people}, \emph{devices}, and
\emph{existing software}.

\newcommand{\ctext}[1]{\begin{tabular}{c}#1\end{tabular}}

\begin{figure}[htb!]
  \centering

\begin{tikzpicture}[>=stealth'] 

  \tikzstyle{every label}=[blue!40!black,font=\bfseries]
  \tikzstyle{every node}=[font=\footnotesize] 
  \tikzstyle{every text node part/.style}=[align=center]

  \tikzstyle{rect}=[draw, rectangle, align=center,minimum height=5ex,
    text width=8ex]

\node [font=\bfseries,color=blue!40!black] (why) at (0,5.5) {Why?};
\node [font=\bfseries,color=blue!40!black] (what) at (0,3.2) {What?};
\node [font=\bfseries,color=blue!40!black] (who) at (0,0) {Who?};

\node [cloud,draw,align=center,aspect=3,label={System to be}] (cloud) at (7.0,5.75) {Objectives};

\node [ellipse,draw,dashed,inner sep=0.5,label={System as is}] (core) at (2.4,5.55) {\begin{tabular}{c}Problems,\\Opportunities,\\Knowledge\end{tabular}};

\node [ellipse,inner sep=0.5,draw] (core) at (7.0,3.2) {\ctext{Services,\\Constraints,\\Assumptions}};

\path (0,0)
node [rect,fill=blue!20!gray!40,draw=none] (S) at (4.5,0) {Software to be}
node [rect,fill=blue!20!gray!40,draw=blue!50!gray!60,outer sep=2] (P) at (6.5,0) {People}
node [rect,fill=blue!20!gray!40,draw=blue!50!gray!60,outer sep=2] (D) at (8.0,0) {Devices}
node [rect,fill=blue!20!gray!40,draw=blue!50!gray!60,outer sep=2] (E) at (9.5,0) {Existing software};

\node [fill=green!5!gray!40,text width=50ex,rounded corners, align=right] (sat) at (7.2,4.5) {Satisfy};

\node [fill=green!5!gray!40,align=right, text width=50ex,rounded corners] (A) at (7.2,1.9) {Assigned to};

\draw [->,thin] (core)--(S);
\draw [->,thin,shorten >=-2] (core) to[](P);
\draw [->,thin,shorten >=-2] (core)--(D);
\draw [->,thin,shorten >=-2] (core)--(E);
\draw [->,thin] (core)--(cloud);

\node [draw=none,above=0.3 of E] (lbl) {\ctext{Environment}};

\begin{pgfonlayer}{background}
    \filldraw [line width=2,join=round,black!10]
      (lbl.north-| E.east)  rectangle (P.south-| P.west);
\end{pgfonlayer}

\end{tikzpicture}

\caption{Current RE approaches, conceptually
  \protect\cite{vanLamsweerde:requirements:2009}.}
\label{fig:RE-conceptual-model}
\end{figure}
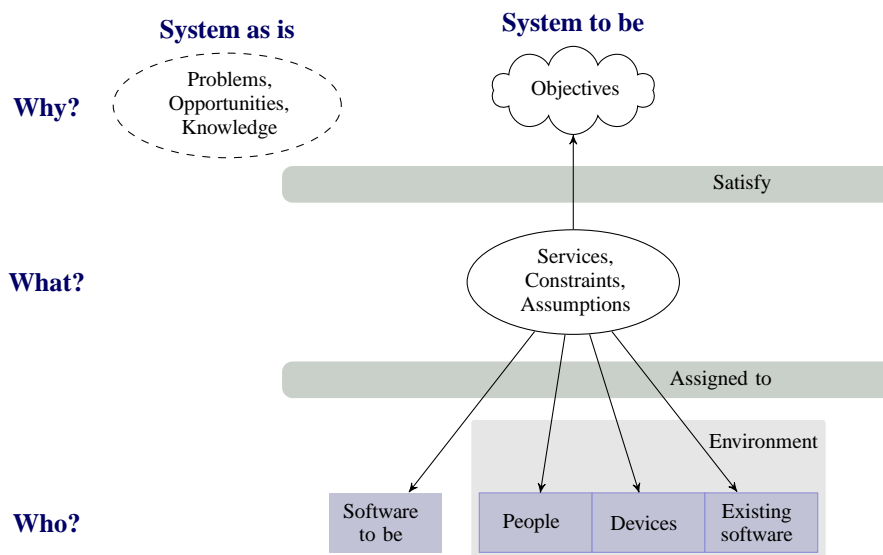

Even more generally, given a set of requirements, the essential idea
is to come up with the specification of a \emph{machine}
(equivalently, the \fsl{software to be} above) that along with
reasonable domain assumptions satisfies the requirements of the
stakeholders \cite{zave:dark-corners:1997}.  The \emph{system as is}
is the system as it exists without the machine and helps us understand
the environment in which the machine will be introduced.  The
\fsl{system to be} is what the system will be when the machine is
introduced.  If all goes well in the \fsl{system to be}, the
stakeholders' requirements are satisfied.

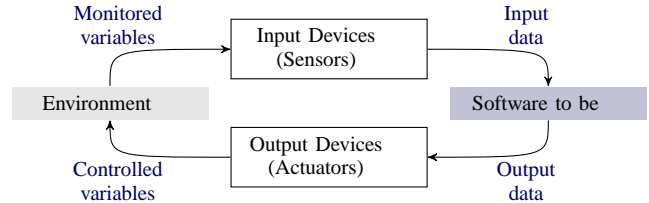
\begin{figure}[htb!]
\centering
\begin{tikzpicture}[node distance=1.3cm,>=stealth',bend angle=45,auto]

  \tikzstyle{box}=[thick,draw=none,text width=15ex,align=center]
  \tikzstyle{abox}=[box,fill=blue!20!gray!40]
  \tikzstyle{ebox}=[box,fill=black!10]
  \tikzstyle{IO box}=[box,draw=black,thin,fill=white,text=black]

  \tikzstyle{edge_label}=[blue!40!black,draw=none,text width=10ex,align=center]
  \tikzstyle{every node}=[font=\footnotesize] 
  \tikzstyle{every text node part/.style}=[align=center]

\matrix (machine_envt) [draw=none,row sep=1mm, column sep=3mm] 
 {
    \node [draw=none] (empty-TL) {}; &
    \node [IO box] (input) {Input Devices\\(Sensors)}; &
    \node [draw=none] (empty-TR) {};\\
    \node [ebox] (environment) {Environment}; &&
    \node [abox] (software) {Software to be}; \\
    \node [draw=none] (empty-BL) {}; &
    \node [IO box] (output) {Output Devices\\(Actuators)}; &
    \node [draw=none] (empty-BR) {};\\
};
\draw [->] (environment) ..  controls (empty-TL) ..  node
      [edge_label,above] {Monitored\\variables}
      (input);
\draw [->] (input) ..  controls (empty-TR) ..  node
      [edge_label,above] {Input\\data}
      (software);
\draw [->] (output) ..  controls (empty-BL) ..  node
      [edge_label,below] {Controlled\\variables}
      (environment);
\draw [->] (software) ..  controls (empty-BR) ..  node
      [edge_label,below] {Output\\data}
      (output);
\end{tikzpicture}
\caption{The machine-environment interface
  \protect\cite{vanLamsweerde:requirements:2009}.}
\label{fig:machine-environment}
\end{figure}

It is important to understand the nature of the machine.
Figure~\ref{fig:machine-environment} (from
\cite{vanLamsweerde:requirements:2009} but based on
\cite{parnas:system-model:1995}) illustrates an operational
conceptualization of the machine-environment interface.  Here, the
\emph{software to be} represents the machine.  A machine is a
\emph{controller} that maps inputs from its environment (by monitoring
certain variables) to outputs or effects in the environment (by
setting certain variables).  In other words, it processes inputs to
produce outputs.  To its users (in the environment), a machine
provides computational services (functionality).

In its very conception, a machine corresponds to a single locus of
control.  Indeed, it is common in software engineering to talk of
machines as acting in pursuit of goals
\cite{vanLamsweerde:requirements:2009}.  Traditional SE is
machine-oriented---it concerns itself with the specification of a
machine (even if implemented via distributed components) that would
meet stakeholder requirements.  Thus traditional SE reflects a
\emph{conceptually centralized} perspective, namely, of the
stakeholders, considered collectively.

The above account of machine-orientation, taken from leading writings
on SE, is agnostic to particular software methodologies.
Machine-orientation manifests itself in specific modeling notations
and methodologies, some of which have been highly influential in SE.
Many existing requirements-based approaches, although differing in
their details, instantiate the same concepts at their core: machine,
environment, \fsl{system as is}, and \fsl{system to be}.  Tropos
\cite{bresciani:tropos:2004} and i* \cite{yu:i-star:1997} refer to a
model of the \fsl{system as is} and the \fsl{system to be} as the
early and late requirements models, respectively.  In their
terminology, the environment is a set of \emph{actors}; the machine
itself is the \emph{system actor}.  For example, Tropos would create a
system actor for an entire healthcare system.  This actor would
capture (a consistent subset of) the goals of all stakeholders,
thereby functioning as a logically centralized machine.  Following
KAOS \cite{dardenne:goals:1993}, one would create a set of
\emph{agents} with designer-assigned individual goals.  However, the
set of agents is conceptually a single machine because there is a
single locus of design.  This follows from the fact that goals are
assigned to agents and, therefore, at a high-level the agents have
already been specified.  Indeed, referring specifically to KAOS (and
to Feather's work \shortcite{feather:composite-systems:1987}, which
provides the conceptual basis for KAOS), Zave and Jackson
\shortcite{zave:dark-corners:1997} observe that even though KAOS
supports multiple agents, it is only a refinement of their more basic
single-machine framework.

Let us apply the above conceptual model of systems to the healthcare
scenario.  Imagine that a healthcare machine were built to a specified
set of stakeholder requirements.  The machine would control various
medical devices and other equipment.  Users such as consumers,
providers, MCOs, and so via on would interface with the machine via a
Web interface.  Consider the example of scheduling an appointment with
a physician.  The physician would have configured his or her
preferences in the machine: his or her daily schedule, how long a time
period to allot each patient, and so on.  Those are the physician's
inputs to the machine, which the machine uses in scheduling patient
appointments.  The patient's interface would display the available
slots and enable him or her to choose from among those slots.  The
machine records each selection, and produces as output a confirmation
of the appointment, perhaps also by email.  Further, the machine grays
out the slot for future scheduling.  The machine additionally supports
the processing of claims and payments among hospitals and insurers.
The machine would encode the insurance company's process of queuing
claims above \$10,000 separately for detailed examination.  The
insurance company through its interface can obtain the claims from the
appropriate queues.  The machine would implement the appropriate
access control so that users can access the appropriate functionality.

Traditional SE, in spite of its conceptually centralized perspective,
may be applied to yield a physically distributed machine.  Web
applications suggest physical distribution and applying KAOS would
result in a physically distributed set of agents.  In general,
techniques from software architecture may be applied to internally
decompose a machine into distributed components.  Rapanotti {\etal}
\shortcite{rapanotti:problem-frames-decomposition:2004} employ such an
approach in the decomposition of machines in the problem frames
approach.

\section{IOSE Concepts}
\label{sec:IOSE}
We define a \emph{protocol} of a sociotechnical system as a
specification of how interactions among its roles would proceed.  In
business settings, the protocol can take the form of a business
contract that principals would enter into upon adopting different
roles.  For example, in Texas, if MedCo wishes to play the role of
\fsc{mco}, it must enter into a \emph{Uniform Managed Care} contract
with the state health commission (which alone plays the
\fsc{regulatory body} role) \cite{texas:managed-healthcare:2012}.

At any point during the interaction, the set of social expectations
represents the \emph{social state} of the system.  A protocol serves
three purposes.  One, it makes public the social expectations of the
participants while giving them the flexibility to follow their
individual objectives.  For example, public funding agencies may
expect that MCOs, upon notification, refund payments made in error
within 30 days.  Two, it identifies which participant is accountable
to which participant for what expectation.  For instance, if MedCo,
who plays \fsc{mco}, does not refund an erroneous payment in time, the
funding agency may hold MedCo accountable.  Three, it frees
participants to implement their information systems as they please as
long as they satisfy the given protocol.  For example, MedCo might
employ any representation it chooses and may apply its policies in
deciding whether to provide a refund early or late within the 30-day
window.

\begin{figure}[htb!]
\centering
\begin{tikzpicture}[node distance=1.3cm,>=stealth',bend angle=45,auto]

  \tikzstyle{box}=[thick,draw=none,text width=11ex,align=center]
  \tikzstyle{abox}=[box,fill=green!10!blue!30!gray!60]
  \tikzstyle{ebox}=[box,fill=black!10]
  \tikzstyle{rbox}=[box,fill=green!10!blue!30!gray!40]

  \tikzstyle{IO box}=[box,draw=black,thin,fill=white,text=black]

  \tikzstyle{edge_label}=[blue!40!black,draw=none,text width=10ex,align=center]
  \tikzstyle{every node}=[font=\footnotesize] 
  \tikzstyle{every text node part/.style}=[align=center]

\matrix (machine_envt) [draw=none,row sep=1mm, column sep=3mm] 
 {
    \node [ebox] (environment-L) {Environment};
     &
     &&
    \node [ebox] (environment-R) {Environment};\\
    &\node [draw=none,ellipse,minimum height=5ex] (empty-top-L) {};
    &\node [draw=none,ellipse,minimum height=5ex] (empty-top-R) {};\\
    \node [draw=none] (empty-principal-L) {}; &
    \node [abox] (principal-L) {Principal}; &
    \node [abox] (principal-R) {Principal}; &
    \node [draw=none] (empty-principal-R) {};
    \\
    \node [draw=none,ellipse,minimum height=2ex] {};\\
    \node [draw=none] (empty-ML) {}; &
    \node [rbox] (role-L) {Role}; &
    \node [rbox] (role-R) {Role}; &
    \node [draw=none] (empty-MR) {};\\
    \node [draw=none,ellipse,minimum height=3ex] {};\\
    & \node [draw=none,minimum height=1.1ex] (empty-BCL) {}; &
    \node [draw=none,minimum height=1.1ex] (empty-BCR) {};\\
};

\node [fill=none,fit=(empty-top-L) (empty-top-R),minimum
  height=1.2ex]
(Sociotechnical system) {Sociotechnical system as Protocol};

\node [fill=none,fit=(empty-BCL) (empty-BCR),minimum height=1.2ex]
(protocol) {\\Meanings};

\draw [<->,dashed,draw] (principal-L.south) ..  controls (protocol.south) ..  (principal-R.south);

\draw [<->,draw] (environment-L.south) ..  controls (empty-principal-L) ..  (principal-L.west);
\draw [<->,draw] (environment-R) ..  controls (empty-principal-R) ..  (principal-R);

\begin{pgfonlayer}{background}
    \filldraw [line width=2mm,join=round,blue!20!gray!20]
      (empty-top-L.north-| principal-R.east)  rectangle (protocol.south-| principal-L.west);
\end{pgfonlayer}


\end{tikzpicture}
\caption{The IOSE approach schematically.}
\label{fig:system}
\end{figure}
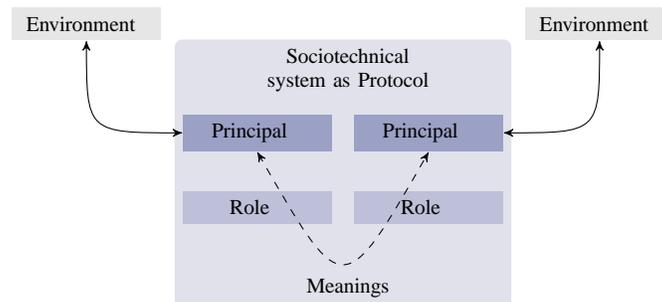

Accordingly, the objective of IOSE is to specify a
\emph{sociotechnical system} as a protocol that defines two or more
\emph{roles}.  Figure~\ref{fig:system} describes how IOSE applies.
For the sociotechnical system to be instantiated, principals adopt
roles in the protocol.  Each principal continues to interact with its
environment even after it joins the system.  The principals
communicate with each other within the scope of the system, which
means that they are subject to the meanings defined in the protocol
that specifies the system.  In engineering terms, the main artifact
produced by IOSE is the protocol which, through its roles, itself
serves as a requirement for the principals who would adopt those
roles.

Whereas traditional SE seeks to specify a machine, IOSE seeks to
specify a protocol.  Whereas a machine captures a single locus of
control, IOSE naturally supports multiple perspectives as roles in a
protocol.  In IOSE, each principal, who adopts one or more roles, is a
locus of control---a reasoner and a sender and receiver of
communications.

\subsection{Commitment Protocols as an IOSE Approach}
We introduce commitment protocols \cite{yolum:flexible-protocols:2002}
as an example IOSE approach.  A (social) commitment captures an
elementary business relationship between a debtor and a creditor
\cite{singh:ontology-commitments:1999}.  Specifically, the commitment
$\C(x,y,r,u)$ says that the debtor $x$ commits to the creditor $y$
that if the antecedent $r$ comes to hold, then it will bring the
consequent $u$.  The elementary operations on commitments are
\fsl{Create}, \fsl{Release}, \fsl{Cancel}, \fsl{Delegate}, and
\fsl{Assign}.  The social nature of commitments owes to the fact that
they progress due to interactions among principals, not due to the
internal reasoning of any principal.  Commitment protocols exploit
this connection by specifying the meanings of \emph{messages} in terms
of how they affect commitments, thus enabling principals to interact
flexibly.  Other abstractions could potentially be used in addition to
commitments but, for simplicity, we confine the present discussion to
commitments.

Table~\ref{table:scheduling-protocol} shows a partial appointment
scheduling protocol.  The protocol involves two roles, \fsc{phy} (for
physician) and \fsc{pat} (for patient).  Assume that via the enactment
of some other protocol, (any physician playing) \fsc{phy} is committed
to (any patient playing) \fsc{pat} to provide the latter with a list
of available appointment slots upon the latter's request for
appointment (this commitment, for instance, could be set up when a
patient registers with a physician for the first time).  We represent
this commitment as
$\C(\fsc{phy},\fsc{pat},\msf{requestAppointment}(\fsc{pat},
\fsc{phy}),\msf{availableSlots}(\fsc{phy}, \fsc{pat}, \vec{s}))$.  The
\fsl{availableSlots} message from \fsc{phy} to \fsc{pat} conveys the
list of available slots to the \fsc{pat}: it means that \fsc{phy}
commits to \fsc{pat} that if \fsc{pat} commits to show up for one of
those slots, then \fsc{phy} will commit to showing up for that slot as
well.  The \fsl{select} message from \fsc{pat} to \fsc{phy} commits
\fsc{pat} to a selected slot.  The \fsl{confirmSlot} message from
\fsc{phy} to \fsc{pat} for a particular slot commits \fsc{phy} to the
slot.  A complete and effective meeting scheduling protocol would
additionally include messages that deal with meeting cancellation and
rescheduling.

\begin{table}[htb!]
\centering
\tbl{A partial appointment scheduling protocol.\label{table:scheduling-protocol}}{%
\begin{tabular}{lp{6.75cm}}\toprule

  \fbf{Message} & \fbf{Meaning} \\ \midrule

  \fsl{availableSlots}(\fsc{phy}, \fsc{pat}, $\vec{s}$) &
 $\C(\fsc{phy},\fsc{pat},\exists s\in \vec{s}:\C(\fsc{pat},\fsc{phy},\top,\msf{showUp}(\fsc{pat},s)), $\\
& \quad$\C(\fsc{phy},\fsc{pat},\top,\msf{showUp}(\fsc{phy},s)))$\\

  \fsl{selectSlot}(\fsc{pat}, \fsc{phy}, $s$) & $\C(\fsc{pat},\fsc{phy},\top,\msf{showUp}(\fsc{pat},s))$\\

  \fsl{confirmSlot}(\fsc{phy}, \fsc{pat}, $s$) &  $\C(\fsc{phy},\fsc{pat},\top,\msf{showUp}(\fsc{phy},s))$\\
\bottomrule
\end{tabular}
}
\end{table}

\begin{table}[htb!]
\centering
\tbl{Commitments involved in the appointment scheduling protocol.\label{table:scheduling-protocol-commitments}}{%
\begin{tabular}{lp{11.75cm}}\toprule
  \fbf{Label} & \fbf{Commitment} \\ \midrule

$c_0$ & $\C(\mathit{Alessia},\mathit{Bianca},\msf{requestAppointment}(\mathit{Bianca},
\mathit{Alessia}),\msf{availableSlots}(\mathit{Alessia}, \mathit{Bianca}, \vec{s}))$\\
$c_{1}$ & $\C(\mathit{Alessia},\mathit{Bianca},\top,\msf{availableSlots}(\mathit{Alessia}, \mathit{Bianca}, \vec{s}))$\\
$c_2$ &
$\C(\mathit{Alessia},\mathit{Bianca},\exists s\in \{1400,1600\}:\C(\mathit{Bianca},\mathit{Alessia},\top,\msf{showUp}(\mathit{Bianca},s)), $\\

& \quad$\C(\mathit{Alessia},\mathit{Bianca},\top,\msf{showUp}(\mathit{Alessia},s)))$\\
$c_{3}$ & $\C(\mathit{Alessia},\mathit{Bianca},\top, \C(\mathit{Alessia},\mathit{Bianca},\top,\msf{showUp}(\mathit{Alessia},1400)))$ \\
$c_4$ & $\C(\mathit{Bianca},\mathit{Alessia},\top,\msf{showUp}(\mathit{Bianca},1400))$\\
$c_5$ & $\C(\mathit{Alessia},\mathit{Bianca},\top,\msf{showUp}(\mathit{Alessia},1400))$\\
\bottomrule
\end{tabular}
}
\end{table}

\begin{figure}[htb!]
\centering
\begin{tikzpicture}[node distance=170,>=stealth']

  \tikzstyle{box}=[thick,draw=none,text width=7ex,minimum height=7ex,align=center]
  \tikzstyle{abox}=[box,fill=green!10!blue!30!gray!60]

  \tikzstyle{edge_label}=[blue!40!black,auto=true,font=\footnotesize]
  \tikzstyle{edge_label_in}=[edge_label,fill=white,auto=false]

  \matrix (soc-1) [draw=blue!20!black,row sep=25, column sep=0]
    {
      \node [abox] (Alessia-1) {Alessia\\\fsc{mi}};\\
      \node [abox] (Bianca-1) {Bianca\\\fsc{mp}};\\
    };
    \matrix (soc-2) [draw=blue!20!black,row sep=25, column sep=0,right of=soc-1]
     {
       \node [abox] (Alessia-2) {Alessia\\\fsc{mi}};\\
       \node [abox] (Bianca-2) {Bianca\\\fsc{mp}};\\
     };

    \matrix (soc-3) [draw=blue!20!black,row sep=25, column sep=0,right of=soc-2]
     {
       \node [abox] (Alessia-3) {Alessia\\\fsc{mi}};\\
       \node [abox] (Bianca-3) {Bianca\\\fsc{mp}};\\
     };

    \matrix (soc-4) [draw=blue!20!black,row sep=25, column sep=0,below of=soc-3,left of=soc-3,node distance=90]
     {
       \node [abox] (Alessia-4) {Alessia\\\fsc{mi}};\\
       \node [abox] (Bianca-4) {Bianca\\\fsc{mp}};\\
     };

    \matrix (soc-5) [draw=blue!20!black,row sep=25, column sep=0,left of=soc-4]
     {
       \node [abox] (Alessia-5) {Alessia\\\fsc{mi}};\\
       \node [abox] (Bianca-5) {Bianca\\\fsc{mp}};\\
     };

\draw [->,draw] (Alessia-1.south) to node [edge_label_in] {$\C_0$} (Bianca-1.north);

\draw [->,draw] (Alessia-2.south) to node [edge_label_in] {$\C_1$} (Bianca-2.north);

\draw [->,draw] (Alessia-3.south) to node [edge_label_in] {$\C_2$} (Bianca-3.north);

\draw [->,draw] (Alessia-4.240) to node [edge_label_in] {$\C_3$} (Bianca-4.120);
\draw [->,draw] (Bianca-4.60) to node [edge_label_in] {$\C_4$} (Alessia-4.300);

\draw [->,draw] (Alessia-5.240) to node [edge_label_in] {$\C_5$} (Bianca-5.120);
\draw [->,draw] (Bianca-5.60) to node [edge_label_in] {$\C_4$} (Alessia-5.300);

\draw [->,draw,very thick, blue] (soc-1.east) to node [pos=0.5,edge_label_in] {\fsl{requestAppointment}()} (soc-2.west);

\draw [->,draw,very thick, blue] (soc-2.east) to node [pos=0.5,edge_label_in] {\fsl{availableSlots}(1400,1600)} (soc-3.west);

\draw [->,draw,very thick, blue] (soc-3.south) to node [pos=0.5,edge_label_in,xshift=10] {\fsl{selectSlot}(1400)} (soc-4.east);

\draw [->,draw,very thick, blue] (soc-4.west) to node [pos=0.5,edge_label_in] {\fsl{confirmSlot}(1400)} (soc-5.east);


\end{tikzpicture}
\caption{Progression of the social state during an enactment of the
  appointment scheduling protocol.}
\label{fig:MS-progression}
\end{figure}
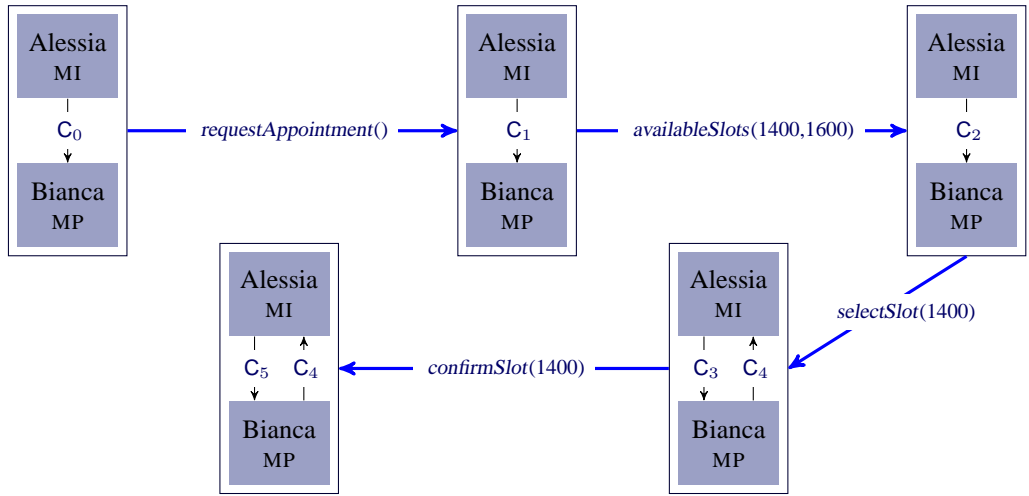

Figure~\ref{fig:MS-progression} depicts the progression of social
state according to the meanings in
Table~\ref{table:scheduling-protocol} and using the commitments
introduced in Table~\ref{table:scheduling-protocol-commitments}.
Alessia plays \fsc{phy} and Bianca plays \fsc{pat}.  Upon Bianca
sending the \fsl{requestAppointment} message to Alessia, the
antecedent of $c_0$ is satisfied.  This makes Alessia unconditionally
committed to sending her the available slots ($c_1$).  When Alessia
sends the list of available slots (at 1400 and 1600 hours), she
discharges $c_1$ and creates $c_2$.  Bianca communicates her selection
of 1400, which makes Alessia unconditionally committed to confirming
it; in other words, Bianca creates $c_4$, which creates $c_3$ (the
unconditional version of $c_2$).  Finally, Alessia confirms the
selected slot, thus creating $c_5$ and discharging $c_3$.  In the
final state of Figure~\ref{fig:MS-progression}, both Alessia and
Bianca are committed to each other for showing up at the selected
time.

Protocols for handling claims between providers and MCO would be more
complex than those for appointment scheduling.  Desai {\etal}
\shortcite{desai:amoeba:2010} specify protocols for automobile
insurance and show how to compose protocols.  We focus on appointment
scheduling because (1) it illustrates the essential concepts of IOSE,
and (2) it bears similarity to meeting scheduling, an application that
has been studied extensively in the literature, and which we discuss
in detail in Section~\ref{sec:prominent-approaches}.

\section{Evaluating Machine-Orientation and IOSE}
\label{sec:eval-machine-IOSE}
\begin{figure}[htb!]
  \tikzstyle{box}=[thin,draw,align=center,anchor=center,minimum height=20,inner sep=2pt,font=\large]

  \tikzstyle{activity}=[box,fill=blue!20!gray!20,draw=none,minimum
    height=6ex,rounded corners]

  \tikzstyle{agent}=[box,draw=none,circle,fill=green!10!blue!30!gray!60,minimum width=6]

  \tikzstyle{scloud}=[shape=cloud,align=center,thick,aspect=4,fill=green!10!blue!30!gray!40]

  \tikzstyle{annotation}=[blue!40!black,draw=none]
  \tikzstyle{edge_label}=[annotation,midway,anchor=south,align=center]

  \tikzstyle{every text node part/.style}=[align=center]

  \tikzstyle{every even column/.style}=[anchor=base east]
  \tikzstyle{every odd column/.style}=[anchor=base east]

\centering
\begin{subfigure}[b]{0.3\textwidth}
\centering
\begin{tikzpicture}[scale=1.1]

\matrix (trust) [draw=none,matrix of nodes, row sep=10,column sep=0] {
%
  \node[agent] (xx) {\textsf{X}}; &[35]&[35]
  \node[agent] (yy) {\textsf{Y}}; \\
  \node[draw=none] (empty-TL) {};&&
  \node[draw=none] (empty-TR) {};\\
  \node[draw=none] (empty-ML) {};&&
  \node[draw=none] (empty-MR) {};\\
  \node[draw=none] (empty-BL) {};&&
  \node[draw=none] (empty-BR) {};\\
};

\draw let \p1 = (xx.west),
          \p2 = (yy.east),
	  \p3 = (empty-BL.south west),
	  \p4 = (empty-ML)
      in
        node (BL) at (\x1+0,\y3) {}
        node (BR) at (\x2-0,\y3) {}
        node (TL) at (\x1+0,\y4-10) {}
        node (TR) at (\x2-0,\y4-10) {};

\draw node [activity,fit= (BL) (BR) (TL) (TR) (empty-BL.south) (empty-BR.south),inner sep=0,minimum height=40] (communication) {};

\draw let \p1 = (communication.west),
          \p2 = (communication.east),
	  \p3 = (communication.north)
      in
        node [font=\large,draw=none,align=center,minimum height=40,anchor=north] at (\x3,\y3-0)
	{\textsf{Communication}\\{\small \textsf{Ignores Social State}}};

\draw [<->,thick,dashed,red] (yy.south) |-(yy.south|-communication.center)-| (xx.south);

\end{tikzpicture}
\caption{Social opacity.}
\label{fig:old}
\end{subfigure}
\quad				
\begin{subfigure}[b]{0.3\textwidth}
\centering
\begin{tikzpicture}[scale=1.1]

\matrix (trust) [draw=none,matrix of nodes, row sep=10,column sep=0] {
  \node[agent] (xx) {\textsf{X}}; &[35]&[35]
  \node[agent] (yy) {\textsf{Y}}; \\
  \node[draw=none] (empty-TL) {};&
  \node[draw=none] (empty-TR) {};\\
  \node[draw=none] (empty-ML) {};&
  \node[draw=none] (empty-MR) {};\\
  \node[draw=none] (empty-BL) {};&&
  \node[draw=none] (empty-BR) {};\\
%
};

\draw let \p1 = (xx.west),
          \p2 = (yy.east),
	  \p3 = (empty-BL.south west),
	  \p4 = (empty-ML)
      in
        node (BL) at (\x1+0,\y3) {}
        node (BR) at (\x2-0,\y3) {}
        node (TL) at (\x1+0,\y4-10) {}
        node (TR) at (\x2-0,\y4-10) {};

\draw node [activity,fit= (BL) (BR) (TL) (TR) (empty-BL.south) (empty-BR.south),inner sep=0,minimum height=40] (communication) {};

\draw let \p1 = (communication.west),
          \p2 = (communication.east),
	  \p3 = (communication.north)
      in
        node [font=\large,draw=none,align=center,minimum height=40,anchor=north] at (\x3,\y3-0)
	{\textsf{Machine}\\{\small \textsf{Controls Social State}}};

\draw [->,thick, blue] (xx.south) to node [edge_label,right] {API} (xx.south|-communication.north);

\draw [->,thick, blue] (yy.south) to node [edge_label,left] {API} (yy.south|-communication.north);

\end{tikzpicture}
\caption{Social control.}
\label{fig:machine}
\end{subfigure}
\quad				
\begin{subfigure}[b]{0.3\textwidth} 
\centering
\begin{tikzpicture}[scale=1.1]

\matrix (trust) [draw=none,matrix of nodes, row sep=10,column sep=0] {
  \node[draw=none]  {}; &
  \node [scloud] (social) {\textsf{Social State}}; &
  \node[draw=none] {}; \\
  \node[agent] (xx) {\textsf{X}}; &&
  \node[agent] (yy) {\textsf{Y}}; \\
  \node[draw=none] (empty-TL) {};&[70]
  \node[draw=none] (empty-TR) {};\\
  \node[draw=none] (empty-ML) {};&&
  \node[draw=none] (empty-MR) {};\\
  \node[draw=none] (empty-BL) {};&&
  \node[draw=none] (empty-BR) {};\\
};

\draw let \p1 = (xx.west),
          \p2 = (yy.east),
	  \p3 = (empty-BL.south west),
	  \p4 = (empty-ML)
      in
        node (BL) at (\x1+0,\y3) {}
        node (BR) at (\x2-0,\y3) {}
        node (TL) at (\x1+0,\y4-10) {}
        node (TR) at (\x2-0,\y4-10) {};

\draw node [activity,fit= (BL) (BR) (TL) (TR) (empty-BL.south) (empty-BR.south),inner sep=0,minimum height=40] (communication) {};

\draw let \p1 = (communication.west),
          \p2 = (communication.east),
	  \p3 = (communication.north)
      in
        node [font=\large,draw=none,align=center,minimum height=40,anchor=north] at (\x3,\y3-0)
	{\textsf{Communication}\\{\small \textsf{Realizes the Protocol}}};

\draw [<->,thick,blue] (xx) to node [edge_label,anchor=north] {Protocol} (yy);

\draw [<->,thick,dashed,red] (yy.south) |-(yy.south|-communication.center)-| (xx.south);

\draw [very thick,red!20!green!90,dotted] (social.south) |-(social.south|-xx.east);

\end{tikzpicture}
\caption{IOSE.}
\label{fig:IOSE}
\end{subfigure}
\caption{A historical perspective on the placement and computation of
  social state.}
\label{fig:system-types}
\end{figure}
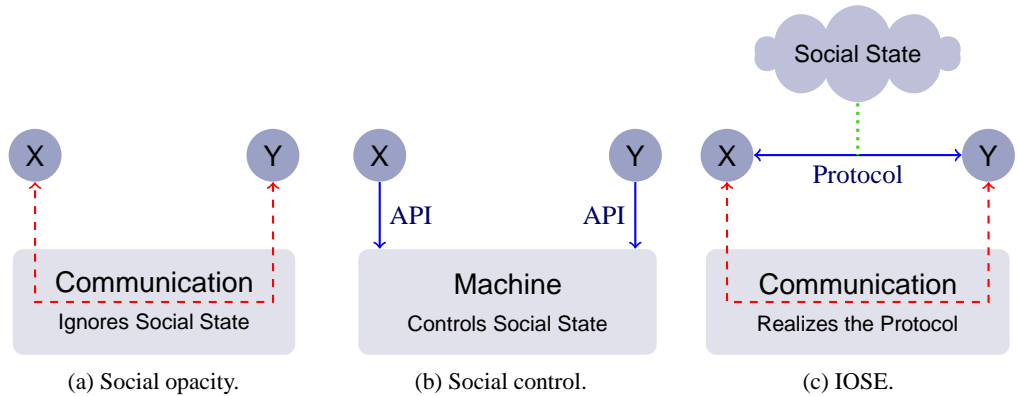

Figure~\ref{fig:system-types} describes three takes on a
sociotechnical system.  Figure~\ref{fig:old} shows the most
traditional setting, which even predates IT.  Here, one can imagine a
sociotechnical system as realized by two or more principals, each with
its own ledgers, and interacting via the postal service or foot
messenger.  The principals are autonomous.  The same situation holds
in the case of digital communication, where we can think of the
messaging system as providing the necessary message delivery
functionality.  In this case too, the messages the principals send and
receive are opaque to the communication network.  In either case,
there is no computational support for the social state.  Indeed, there
is no computational support for anything beyond message transport and,
possibly operational constraints such as message ordering.  Each
principal maintains its expectations and commitments in its ledgers
and acts according to its local policies.  For example, a physician
may send a request to a laboratory to find a patient's cholesterol and
the laboratory may send back the results along with an invoice.

Figure~\ref{fig:machine} shows a more sophisticated, but traditional,
approach wherein a machine mediates all interactions among the
principals.  The machine offers an API to the principals through which
they can request changes in the social state.  Such conceptually
centralized machines are not common for healthcare today, so we use
eBay as an example and then return to healthcare.  The auctioneer eBay
offers a machine for conducting auctions that provides an API
(requiring little more than a web browser) through which sellers can
list items for sale and buyers can bid for listed items.  The machine
determines the social state: whether something is on sale, its reserve
price, the time an auction closes, whether a bid is legitimate, the
winning bid, and so on.  The principals can represent their local view
in their ledgers but what matters is the eBay machine's view.  The
machine happens to usually respect the requests of buyers and sellers
but it does not need to.

Returning to the healthcare machine of
Section~\ref{sec:machine-orientation}, the patient may attempt to
schedule an appointment with the physician, but the machine determines
whether the appointment holds, i.e., whether the patient and physician
are each committed to showing up at the appointed time.  The patient
and the physician may maintain their local views of the social state
but what the machine maintains is definitive.  The same situation
arises between the hospital and MedCo regarding their payments.  In
this setting, the only way the parties can interoperate is if they
maintain their local views tightly synchronized with the central
machine.

In contrast, Figure~\ref{fig:IOSE} shows how the principals enact a
protocol with one another.  The enactment takes advantage of
infrastructure such as for communication.  However, the social state
is determined not by the infrastructure but through the enactment of
the protocol.  That is, the protocol specifies how the social state
progresses.  Here too the principals can maintain their individual
ledgers but what the protocol specifies is definitive.  The important
challenge of formalizing protocols to ensure that the local views
remain sufficiently consistent with the protocol-specified ``public''
state without synchronization is out of our present scope.  Chopra and
Singh \shortcite{AAMAS-Alignment-09} present a relevant approach.

\subsection{Criteria for Modeling Sociotechnical Systems}
\label{sec:modeling-criteria}
Let us use the healthcare example of Section~\ref{sec:healthcare} to
motivate the key criteria that any software engineering methodology
for sociotechnical systems must support.

\mypara{Accommodate autonomy}
The autonomy of a principal (a human, organization, or other social
entity) means that it is an independent locus of control with its own
goals and policies, which reflect its business motivations and are
normally private.  Thus even as a participant in a system, each
principal is free in principle to act in pursuit of its own goals,
without any consideration of others.  For example, a hospital may
decide not to entertain patients with a particular insurance provider,
or physicians may increase the panel of patients they treat at the
expense of reducing availability.

Figure~\ref{fig:old} is compatible with autonomy only to the extent
that it does not model social state.  Figure~\ref{fig:machine}
violates autonomy because the machine not only determines the
definitive social state but also controls transitions on it.  The
principals have no direct relationship with each other.  For example,
a bid may be declined and a seller may not be able to change his mind
and accept a bid lower than the reserve price.

Contrast the above with protocols.  A protocol is not a
computationally active entity and cannot provide a service.  A
protocol simply specifies the correctness criteria for interaction
among the participants.  Figure~\ref{fig:IOSE} expresses a protocol
that specifies the appropriate social relationships among potential
principals, capturing their legitimate expectations of each other.
Further, principals would enter into these relationships of their own
volition; no relationship is forced upon a principal, not even to
ensure compliance.

\mypara{Support accountability}
Accountability is the flip side of autonomy.  A participant is free to
act as it pleases; however, it is accountable to those that have
legitimate social expectations of it.  For example, a laboratory may
legitimately expect that the MCO transfer funds for services provided
within a certain time after submission of claims.  This means that the
MCO is accountable to the laboratory for transferring funds in time;
if it does not, it violates the laboratory's expectation.

Figure~\ref{fig:old} provides no support for accountability because it
does not represent the social state.  Figure~\ref{fig:machine} holds
each principal accountable to the central authority and, if it
permits, to the other principals.  For example, a buyer is accountable
to eBay to pay in time.  But if she persuades eBay to grant her an
extension, the seller has no recourse within the eBay system.
Figure~\ref{fig:IOSE} bases accountability on a protocol and thus each
principal is accountable to any other principal in whom it has created
a legitimate expectation.  In particular, through the protocol, each
principal ought to be able to judge the compliance with its
expectations of each principal with whom it interacts.

\mypara{Accommodate loose coupling}
Loose coupling captures the idea that each principal is an independent
locus of design.  It has full control over the design of its own
information systems but no control over the design of other
principals' information systems.  For example, a hospital may store
information about a surgeon's willingness to work the night shift
whereas the insurance company may capture the total charges for
procedures led by that surgeon.  And, hospitals in different
jurisdictions may capture different patient monitoring data during a
surgery.

Figure~\ref{fig:old} is compatible with loose coupling insofar as it
elides the treatment of social state.  In practice, it forces an ad
hoc design of the interaction (e.g., message formats).  Because the
message meanings are hidden, the principals who adapt their
information systems risk failing interoperability.
Figure~\ref{fig:machine} requires that the principals adopt the same
representation of the social state as the machine.  The meanings of
the messages sent over the API could be explicit but they usually are
hardcoded in the machine, and sometimes in the principals' information
systems, if any.  This produces a tighter coupling than desirable
between the principals and the machine (both considered as endpoints):
when the machine changes, the principals must potentially change their
own behaviors and any information systems that support them
accordingly.  By contrast, Figure~\ref{fig:IOSE} employs a protocol
with explicit meanings (not hardcoded in any principals' information
system) for the interactions.  The IOSE approach, in effect, makes
public the extent of the coupling.  The principals may adapt their
information systems without consideration of others as long as they
follow the protocol.

\section{Principles of IOSE}
\label{sec:principles-IOSE}
We introduce the core principles of IOSE.  Specifically, we contrast
how the key principles of SE: modularity and encapsulation
\cite{parnas:decomposition:1972}, separation of concerns
\cite{dijkstra:concerns:1982}, and abstraction are manifested in
current modeling approaches with how they are manifested in IOSE.  We
find that although the intuitions behind the principles hold, the
technical details are completely different and, in many cases,
antithetical to what we encounter in conventional SE.

\subsection{Accountability Modularity: Embedding in the Social World}
Modularity refers to the functional, most commonly hierarchical,
decomposition of systems \cite{parnas:decomposition:1972}.  The
benefit of appropriately modularizing systems is composability.  In
the extreme, modules from different vendors may be composed as needed.

As Figure~\ref{fig:machine-environment} shows, the first-level
application of modularity in SE is the decomposition of the system
into the environment and the machine.  Rapanotti {\etal}
\shortcite{rapanotti:problem-frames-decomposition:2004} apply
architectural patterns to decompose a problem frame into its
constituent problem frames.  KAOS, i*, and Tropos, being
agent-oriented, support finer grained units of modularity.  The agents
therein may have responsibilities or provide services, as captured by
the dependencies among them.

In IOSE, a principal as an autonomous social entity is the natural
unit of modularity.  With autonomy comes accountability, as motivated
by Mamdani and Pitt \shortcite{mamdani:responsible-agents:2000}.
Without the ability to check compliance, though, accountability would
be meaningless.

\emph{Example.}  Although Bianca is free to not show for the
appointment once she commits to a slot, she is accountable for her
commitment.

\emph{Benefit.}  Promotes autonomy by not unduly restricting a
principal's course of action.  Promotes accountability by providing a
basis for ensuring \emph{correctness}: a principal who does not comply
may face sanctions from principals to whom it is accountable for the
given expectation.

Each principal autonomously becomes the debtor of any commitments
\cite{singh:ontology-commitments:1999}.  That is, the debtor must have
initiated an interaction (sent a message) that leads to it being
committed.  In some cases, the message could itself create the
commitment.  In other cases, the debtor may have created some
commitment (as debtor) whereby actions by other parties could lead to
the creation of the given commitment.  Because commitments are created
ultimately due to the communications of the debtor, the debtor is
accountable for them.  Demands placed on a principal other than as the
debtor of a commitment have no bearing on compliance or enforcement.

\subsection{Abstraction: Emphasis on Social Meaning}
Abstraction refers to the level of the concepts used in a
specification.  The ideal abstraction is sufficiently high-level to
hide details and reduce complexity, yet sufficiently low-level to
support drawing the necessary conclusions.  Tropos and i* offer
high-level abstractions such as goals, capabilities, and goal
dependencies.  Sommerville {\etal}
\shortcite{sommerville:responsibility:2009} apply high-level notions
of responsibility and delegation to requirements modeling.  Using
high-level abstractions places requirements away from low-level
notions such as tasks, plans, and state machines.  In contrast, IOSE
emphasizes abstractions that capture the meaning of an interaction.

\mypara{Explicit Social Meaning}
Make all social expectations explicit in the system specification.
The meanings of individual communications must be explicitly
formalized in terms of what they \emph{count as} in the society being
designed.  In general, the meaning involves the creation or other
manipulation of the commitments among the parties involved.

\emph{Example.}  Table~\ref{table:scheduling-protocol} specifies the
message meanings in an appointment scheduling protocol.

\emph{Benefit.}  Explicit social meaning promotes loose coupling.  As
Figure~\ref{fig:MS-progression} illustrates, the meaning captures how
the principals' social state progresses.  The true social state
progresses even if we have no computational representation of it.  But
lacking an explicit meaning, each principal could interpret messages
in incompatible ways.  For example, in the healthcare setting, a
laboratory may interpret the messages with the claim information as
leading to an unconditional commitment from the MCO to honor the
claim.  The MCO, however, may interpret the commitment as being
conditional on the claim being valid.  Such misalignments could have
serious repercussions for the principals (e.g., in producing their
balance sheets) and may lead to legal disputes.  If the principals
negotiate the meanings of the messages and hard-code them in their
information systems, they would produce hidden couplings among
themselves: changes in how one principal handles messages would need
to be propagated to the others.

Additionally, explicit social meaning promotes accountability because
if the meanings are public, then principals can potentially check the
compliance of themselves and others.

\mypara{Solely Social Meaning} A system specification must specify
\emph{only} the possible communications and their meanings and nothing
else.  Further, the meaning must be expressed in terms of social
abstractions such as commitments.  Specifications of any operational
details that have significance at the social level (e.g., a convention
to pay first), must be captured via social abstractions (e.g.,
commitments).  Specifications that capture ordering and occurrence
constraints separately from the meaning violate this principle.  For
example, one could specify in the appointment scheduling protocol that
\fsl{availableSlots} follow \fsl{requestAppointment}.  But here no one
is accountable for the constraint: is the physician at fault for not
delaying sending \fsl{availableSlots} or is the patient at fault for
not sending \fsl{requestAppointment}?  Instead, if this constraint is
necessary as a social requirement, one or more of the principals
should commit to enforcing it, e.g., adopting Marengo {\etal}'s
\shortcite{marengo:regula:2011} approach and placing temporal
constraints in commitments.

Further, specifying the technical infrastructure does not capture a
sociotechnical system.  We might either model the social actor who
provides the infrastructure and engage it via commitments or omit such
technical constraints altogether since they apply at a lower level of
abstraction.

\emph{Example.}  The above ordering constraint can be expressed
\cite{marengo:regula:2011} as $\C(\fsc{phy},\fsc{pat},\top,
\msf{requestAppointment}\cdot\msf{availableSlots})$ where the dot
(`$\cdot$') means \emph{occurs before}.

\emph{Benefit.}  Promotes autonomy and accountability.  No central
controller enforces constraints in a sociotechnical system.  Every
constraint logically ought to be some principal's responsibility.
Expressing a constraint as a hard constraint to be enforced magically
by the environment either under-specifies the functioning of the
system or (in most current thinking) postulates a central entity that
is the sole autonomous principal and can impose its will upon all the
other participants.

\subsection{Separation of Social and Technical Concerns}
Separation of concerns refers to the treatment of each aspect of a
problem independently of \emph{yet} in relation to others.  It refers
to the sorting out of the different threads from what would otherwise
be a tangled mess.  Often, the invocation of this principle is
implicit.  For example, Zave and Jackson's identification of domain
assumptions, machine requirements, and user requirements as the
separate but essential categories for RE is at its heart an
application of this principle.  Dardenne {\etal}
\shortcite{dardenne:goals:1993} and Yu \shortcite{yu:i-star:1997}
express early requirements in the form of goal models, thus separating
the exploration of the problem space from the solution space.
Mylopoulos {\etal}
\shortcite{mylopoulos:nonfunctional-requirements:1992} separate
nonfunctional from functional requirements.  Finkelstein {\etal}
\shortcite{finkelstein:viewpoints:1992} separate concerns explicitly
based on stakeholders, acknowledging the fact that, in general, each
stakeholder has different concerns and may employ different
representations for expressing them.

For sociotechnical systems, we must separate social and technical
considerations.  Principals such as people and organizations must be
distinguished from technical entities such as resources, legacy
systems, software components, devices, communication infrastructure,
and other technical objects in the environment.  This is because
social relationships are meaningful only among principals.  A
principal may only bear a control relationship (e.g., ownership,
invocation, or access) toward a technical entity as may one such
entity toward another.  Only principals are autonomous and
accountable: a patient cannot sue an operating table but can sue a
surgeon or a hospital.

\emph{Example.}  As Figure~\ref{fig:MS-progression} shows, Alessia and
Bianca maintain a social relationship.  However, each of them
maintains and controls an information system, which is not socially
visible.

\emph{Benefit.}  Separating the social and technical entities makes
clear the kinds of relationships that would make sense among them.
Promotes accountability by making clear only principals are
accountable to each other in the social sense.  Enables engineers of
sociotechnical systems to focus solely on the social aspects.

\subsection{Encapsulation: No Principal Internals}
Encapsulation refers to the principle that a module reveal no more
information than is necessary to effectively use it, in particular,
that it reveal no implementation details.

Figure~\ref{fig:machine-environment} highlights the interface between
the machine and the environment but does not bind the machine to any
particular internal implementation.  Zave and Jackson
\shortcite{zave:dark-corners:1997} characterize RE as the process by
which you arrive at the machine-environment interface; anything more
would amount to prematurely determining an implementation.  In i* and
Tropos, dependencies among actors correspond to their interfaces.

The idea of encapsulation, namely, to avoid examining the internals of
a component, remains appropriate in IOSE.  A direct consequence of
this principle is that a sociotechnical system cannot be specified in
terms of mental abstractions such as beliefs, goals, intentions, and
so on---neither of its stakeholders nor of the principals who may
participate in it.  In particular, roles cannot be specified in terms
of mental abstractions.  In IOSE, each role in a protocol refers only
to the social commitments resulting from the communications that a
principal adopting it would be involved in.

\emph{Example.}  Neither the \fsc{phy} role nor the \fsc{pat} role has
goal of scheduling appointment; neither do they have a shared (joint)
goal to schedule appointments.

\emph{Benefit.}  Promotes loose coupling by hiding details not
relevant to the interaction.  Also promotes accountability: a key
reason we cannot use mental concepts to specify a sociotechnical
system is that they make determining compliance impossible
\cite{singh:rethinking:1998}.  As mentioned earlier, accountability is
meaningless if we cannot check compliance.

\subsection{Summary of Principles}
Table~\ref{table:principles} presents the principles and their
benefits at a glance.

\begin{table}[htb!]
\centering
\tbl{IOSE principles for sociotechnical system specification and their benefits.\label{table:principles}}{%
\begin{tabular}{p{2.0cm}p{8cm}p{2.6cm}}  \toprule
\fbf{Principle}   & \fbf{Interpretation} & \fbf{Benefits Promoted}\\ \midrule
Accountability modularity          &  Principals are the basic units of autonomy and, therefore, modularity. Principals are accountable for their communications and the resulting social expectations, e.g., commitments.    &  Autonomy and accountability\\
Explicit social meaning            & The social meaning of communication should be made explicit in system specifications. & Accountability and loose coupling.\\
Solely social meaning              & Specify only communications and their social meaning, not control flow or other kinds of low-level constraints & Autonomy and accountability.\\
Separating social from technical & Social relationships hold among principals, not among principals and technical components and neither among technical components & Modeling perspicuity and accountability. \\
No principal internals           & System specifications should not refer to the internals of principals & Accountability and loose coupling.\\
\bottomrule
\end{tabular}
}
\end{table}

\section{Comparing IOSE with Prominent SE Methodologies}
\label{sec:prominent-approaches}
We now evaluate IOSE with some prominent approaches from SE.  We
choose these either because (1) they are representative of broad
classes of approaches for modeling sociotechnical systems (i*, Tropos,
KAOS), or (2) they give emphasis to interaction and protocols (Gaia
and Choreographies).

\subsection{i* and Tropos}
Lacking a treatment of healthcare in i* and Tropos approaches, we
study Yu's treatment of meeting scheduling in i*
\cite{yu:i-star:1997}, which is similar in spirit to the appointment
scheduling protocol discussed earlier.  Despite its simplicity,
meeting scheduling provides sufficient subtlety to demonstrate various
modeling approaches \cite{vanLamsweerde:meeting-scheduler:1995} and is
extensively used in the literature.  The main requirement in this
scenario is to automate and make efficient some aspects of meeting
scheduling so that the meeting initiator's burden is reduced.

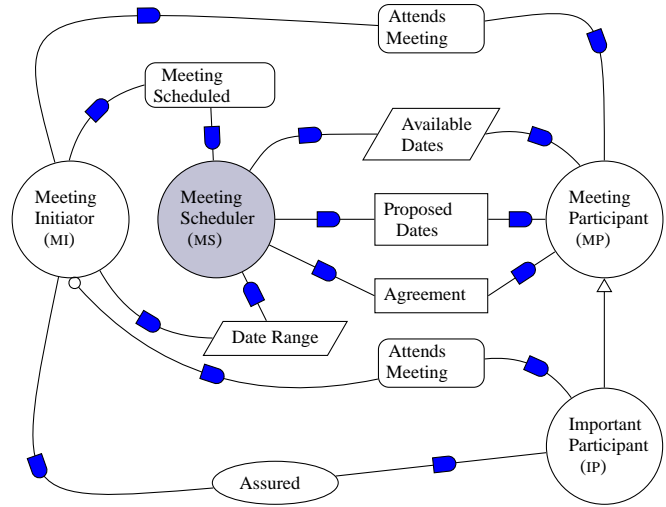
\begin{figure}[!htbp]
\centering
\begin{tikzpicture}

\newcommand{\tropos}{\filldraw[fill=blue,scale around={0.2:(0.0,0.0)},rotate=180] (-0.5,-0.5) arc [start angle=270, end
    angle=90,radius=0.5,draw=none]--(0.5,0.5)--(0.5,-0.5)--cycle;}

\tikzset{myarrow data/.style={
      decoration={
         markings,
         mark=at position #1 with \tropos},
         postaction=decorate}
      }

  \tikzstyle{goal}=[draw,rectangle,rounded corners=4pt,inner sep=2pt,text width=8ex,align=center]
  \tikzstyle{soft}=[draw,ellipse,text width=6ex,align=center]

  \tikzstyle{resource}=[draw,rectangle,text width=8ex,align=center]

  \tikzstyle{actor}=[draw,circle,text width=6ex,align=center]

  \tikzstyle{dependency}=[myarrow data={0.5}]
  \tikzstyle{every node}=[font=\scriptsize] 

\matrix (as-is) [draw=none,row sep=0mm, column sep=2mm,align=center]
 {
   \node [draw=none] (empty-TL) {}; &[0mm]&
   \node [goal,anchor=north] (attends) {Attends\\Meeting}; &[4mm]
   \node [draw=none] (empty-TR) {}; \\
   & \node [goal,anchor=north east,text width=10ex] (scheduled) {Meeting Scheduled}; &
&   \node [draw=none] (empty-two-R) {};\\
   & \node [draw=none] (empty-three-L) {}; &
   \node [trapezium, draw, trapezium left angle=60, trapezium right
     angle=-60,text width=5ex] (available) {Available Dates}; &\\
    \node [actor,anchor=west] (initiator) {Meeting\\Initiator\\(\fsc{mi})}; &
    \node [actor,anchor=east,fill=blue!20!gray!40] (scheduler) {Meeting\\Scheduler\\(\fsc{ms})}; &
    \node [resource] (proposed) {Proposed Dates}; &
    \node [actor] (participant) {Meeting\\Participant\\(\fsc{mp})};\\
    &&
    \node [resource,anchor=south] (agreed) {Agreement};\\[1mm]
    &
   \node [trapezium, draw, trapezium left angle=60, trapezium right
     angle=-60] (range) {Date Range}; &
    \node [goal,anchor=north] (subgoal) {Attends\\Meeting}; \\
    \node [draw=none] (empty-BL) {}; &
    \node [soft,anchor=south] (assured) {Assured}; &&
    \node [actor,anchor=south] (important) {Important\\Participant\\(\fsc{ip})};\\
};

\draw [->,>=open triangle 60] (important) to (participant);

\draw [dependency] (initiator) ..  controls (empty-TL) ..  (attends);
\draw [dependency] (attends) ..  controls (empty-TR) ..  (participant);

\draw [dependency] (initiator) to [bend left] (scheduled.west);
\draw [dependency] (scheduled) to (scheduler);

\draw [dependency] (initiator) to [bend right] (range);
\draw [dependency] (range) to (scheduler);

\draw [dependency] (scheduler) ..  controls (empty-three-L) ..  (available.west);
\draw [dependency] (available.east) to [bend left] (participant);

\draw [dependency] (scheduler) to (proposed);
\draw [dependency] (proposed) to (participant);

\draw [dependency] (scheduler) to (agreed.west);
\draw [dependency] (agreed.east) to (participant);

\draw [dependency] (initiator) ..  controls (empty-BL) ..  (assured);
\draw [dependency] (assured) to (important);

\draw [o-,dependency,bend right] (initiator.south) to (subgoal);
\draw [dependency,bend left] (subgoal) to (important);

\end{tikzpicture}
\caption{The meeting scheduling \fsl{system to be}
  {\protect\cite{yu:i-star:1997}}.  The machine is the Meeting
  Scheduler.}
\label{fig:SD-system-to-be}
\end{figure}

Figure~\ref{fig:SD-system-to-be} shows the \emph{system to be} for a
meeting scheduling system in the i* notation.  The circles represent
the actors in the system: \fsc{meeting initiator} (\fsc{mi}),
\fsc{meeting participant} (\fsc{mp}), and \fsc{meeting scheduler}
(\fsc{ms}).  The directed links between the actors represent
\emph{dependencies}.  For example, \fsc{mi} depends on \fsc{mp} for
achieving the goal that \emph{the participant attend the meeting}
(\fsf{Attends Meeting}).  In the i* terminology,
Figure~\ref{fig:SD-system-to-be} is a \emph{strategic dependency} (SD)
diagram.

According to Yu, in \emph{early requirements engineering}, i* helps
come up with an initial set of requirements.  Specifically, i* helps
identify the goals of the various stakeholders and their dependencies
in achieving those goals.  A new system actor is introduced unto which
some of these goals are \emph{delegated}.  The system actor represents
the machine to be designed to meet stakeholder goals.  This
\emph{goal-modeling} phase helps refine the system actor's goals and
its dependencies with the other actors.  In
Figure~\ref{fig:SD-system-to-be}, the meeting scheduler \fsc{ms} is
the system actor and is responsible for many tasks that \fsc{mi} was
responsible for in the \fsl{system as is} (not shown), such as
obtaining availability information from participants and choosing a
date that suits all participants.

Tropos is an agent-oriented software engineering (AOSE) methodology
that builds on the i* metamodel \cite{bresciani:tropos:2004}.  Tropos
follows i*'s goal modeling phase with an architectural phase that maps
the system actor to new actors that are interconnected through
appropriate data and control dependencies.  The detailed design phase
models each of these actors as one or more belief-desire-intention
agents.  The implementation phase realizes the agents, thereby
completely implementing the system actor.  Both i* and Tropos would
create a system actor for an entire healthcare system.  This actor
would capture (a consistent subset of) the goals of all stakeholders,
thereby functioning as a logically centralized machine.  The comments
below apply equally to i* and Tropos.

\begin{itemize}
\item \emph{Accountability Modularity:} Violated.  There is no
  discrete principal behind the system actor and it is not meaningful
  to talk of its accountability.  For example, \fsc{ms}, the system
  actor in Figure~\ref{fig:SD-system-to-be} is not accountable to the
  stakeholders.  In general, dependencies in i* do not support
  accountability---just because an actor depends on another for a goal
  does not make the latter accountable for it.  For example,
  Figure~\ref{fig:SD-system-to-be} shows that the \fsc{mi} depends on
  the \fsc{mp} for attending the meeting; however it does not
  automatically follow (without reasoning about communication and the
  resulting commitments, as in IOSE) that the \fsc{mp} is accountable
  for showing up.
\item \emph{Explicit Social Meaning:} Partially fulfilled.  The
  purpose behind the notion of dependencies in i* was to capture
  interactor relations in terms of high-level abstractions.  However,
  i* dependencies refer to actors' mental states, and are not social,
  as further explained under encapsulation below.
\item \emph{Solely Social Meaning:} Partially fulfilled.  Dependencies
  are the only interactor abstraction.  However, as we mentioned
  above, i* dependencies are not social abstractions.
\item \emph{Separating Social from Technical:} Partially fulfilled.
  i* treats social and technical actors alike with the same kinds of
  dependencies between any pair of actors.  For example, \fsc{ms} is a
  technical actor whereas \fsc{mp} and \fsc{mi} are stakeholders but
  \fsc{ms}'s dependencies with \fsc{mi} and \fsc{mp} are of the same
  nature as those between \fsc{mi} and \fsc{mp}.  However, to its
  credit, in the early requirements modeling phase, all the actors are
  stakeholders and only in the later phases (starting from late
  requirements modeling) are technical actors introduced.
\item \emph{No Principal Internals:} Violated.  An i* goal dependency between
  two actors $x$ and $y$ for some goal $p$ means that $x$ has a goal
  $p$, and $y$ is \emph{able} to achieve $p$ and in addition
  \emph{intends} to deliver $p$.  For example, as
  Figure~\ref{fig:SD-system-to-be} shows, \fsc{mi} depends on \fsc{ms}
  for its goal \fsf{Meeting Scheduled}.  This dependency means that
  (1) \fsc{mi} has a goal to have the meeting; (2) \fsc{ms} is able to
  schedule the meeting; and (3) \fsc{ms} intends to schedule the
  meeting.  In other words, a goal dependency expresses a shared goal,
  indicating \emph{joint intentionality}, among the actors.  (Soft
  goal, task, and resource dependencies are interpreted analogously.)
  Thus the connection expressed between actors is rooted in their
  internals.  Indeed, Yu states this: ``The Strategic Dependency model
  aims to present a picture of agents by explicitly modeling only
  their external intentional relationships with each other.  The
  semantics of these external relationships, however, are
  characterized in terms of some \emph{presumed internal intentional
    features} of the agent'' \cite[p.~26, emphasis
    added]{yu:i-star:1995}.

\end{itemize}

In summary, the good points about i* and Tropos are that they model
stakeholders and stakeholder relations explicitly as actors and
dependencies.  Their shortcomings include the machine-orientation
underlying the notion of system actor; mentalist dependencies; and
violation of encapsulation.  That is, i* and Tropos can help design a
machine in a cooperative setting, where autonomy is not a
consideration, but do not adequately address the engineering of a
sociotechnical system.

\subsection{KAOS}

Dardenne {\etal}'s \shortcite{dardenne:goals:1993} KAOS resembles
Tropos in its goal-orientation.  KAOS first elicits stakeholders goals
and represents them as AND-OR graphs.  Next, it selects a particular
variant of the graph as the basis for further engineering.  Domain
analysis reveals the possible sets of agents in the \fsl{system to be}
along with their capabilities.  KAOS's strength lies in the
methodological details of deriving operational constraints from goals
and assigning them to particular agents as responsibilities depending
on their capabilities.  For uniformity, we consider examples from van
Lamsweerde {\etal}'s \shortcite{vanLamsweerde:meeting-scheduler:1995}
study of the meeting scheduler.

The notion of \emph{agents} in KAOS is fundamentally different from
that of principals in IOSE.  Agents in KAOS may be social (human) or
technical, e.g., devices and software programs.  Further, an agent is
considered a performer of actions, which are themselves defined in
terms of input, output, precondition, and postcondition.  For example,
van Lamsweerde {\etal} specify a \fsf{scheduler} agent that has the
capability to perform the action \fsf{DetermineSchedule}, whose
preconditions are that the meeting be requested but not scheduled and
whose postconditions are (1) if the meeting is feasible, it is
scheduled and (2) if it is not, the scheduling attempt fails.  Other
agents in van Lamsweerde {\etal}'s meeting scheduling system are
\fsf{participant} and \fsf{initiator}, which are analogously
specified.

In KAOS, a system specification captures adequately refined goals
(operational in nature) and their assignment to agents with the
requisite capabilities.  For example, the goal of notifying
participants is assignable either to the \fsf{initiator} or
\fsf{scheduler} because they are both capable of notifying
participants; the goal of maintaining an up-to-date agenda that the
\fsf{scheduler} could consult is the responsibility of every
\fsf{participant}; and so on.  The idea is that if the agents are
appropriately specified and the goals are appropriately assigned, then
overall system goals would be met.

KAOS specifications capture neither communications among the agents
nor explicit relationships among them.  Instead KAOS conceptualized
agents as monitoring and setting the appropriate system-level (meaning
global) variables and predicates.

KAOS would model the healthcare system via multiple agents but assign
the goals of each agent, thereby creating a conceptually centralized
machine.  The contrast with IOSE principles is stark:

\begin{itemize}
\item \emph{Accountability Modularity:} Violated.  KAOS's notion of
  composite system is exactly that of
  Figure~\ref{fig:machine-environment}: the two components are the
  automated system and the environment
  \cite{vanLamsweerde:meeting-scheduler:1995}.  The users are
  considered part of the environment.  The automated system consists
  of software agents.  Thus KAOS supports the specification of a
  distributed machine (for example, above, we discussed the KAOS
  specification of agents for the meeting scheduler system).  The
  automated system represents the realization of the stakeholder
  requirements; hence it is meaningless to talk its accountability to
  anyone.  We note here that the notion of responsibility in KAOS in
  not a social one: a responsibility is simply an operational
  constraint that is assigned (and therefore, designed) into an agent.

\item \emph{Explicit Social Meaning:} Violated.  KAOS does not model
  communication nor employ any communication-related social
  abstractions.  Further, unlike Tropos, it does not have abstractions
  to capture any interagent relations.  All interaction between agents
  is captured only indirectly---through the setting of the appropriate
  variables and predicates in the environment.

\item \emph{Solely Social Meaning:} Violated.  Since KAOS does not
  support social meaning.

\item \emph{Separating Social from Technical:} Violated.  KAOS models
  a sociotechnical system in a purely technical way.  Neither
  stakeholders nor principals are accommodated in the system model.
  In fact, KAOS does not even model the goals as being of particular
  stakeholders: there is just one goal tree, which is progressively
  refined until the leafs can be assigned to agents.  The names of the
  agents in the meeting scheduling system (\fsf{participant}, and so
  on) may suggest that KAOS accommodates principals.  However, that
  would be misleading: agent specifications together with goal
  assignments are essentially abstract programs.  The entire automated
  system is a collection of programs interacting via variables.

\item \emph{No Principal Internals:} Violated.  The system is
  ultimately a collection of agents whose assigned responsibilities
  determine their implementations.  Since the system's correct
  functioning depends on the goals assigned to its member agents, KAOS
  breaks encapsulation.
\end{itemize}

Like Tropos, KAOS applies in designing a technical system in a
cooperative setting, but its machine-orientation precludes engineering
an open sociotechnical system.

\subsection{Gaia}

Zambonelli {\etal}'s \shortcite{zambonelli:gaia:03} Gaia is one of the
earliest AOSE methodologies.  Later AOSE methodologies bear many
conceptual similarities with Gaia.  Hence, it deserves an in-depth
discussion.

Gaia takes an organization-based approach in which agents may adopt
roles.  Zambonelli {\etal} consider a conference management system in
which agents may adopt the appropriate roles (such as \fsc{reviewer},
\fsc{pc member}, \fsc{author}, and so on).  A role defines the
permissions and responsibilities of a prospective agent.  Permissions
describe what an agent could do with resources in the environment.
Responsibilities are algebraic expressions over the protocols and
internal activities that an agent must perform.  For \fsc{reviewer},
the permissions are \fsf{reads Papers} and \fsf{change ReviewForms}
and the responsibilities are first \fsf{ReceivePaper} (a protocol),
second \fsf{ReviewPaper} (an internal activity), and third
\fsf{SendReview} (a protocol).  Below we discuss Gaia with respect to
the IOSE principles.

\begin{itemize}
\item \emph{Accountability Modularity:} Violated.  In Gaia, technical
  components such as mail clients and active databases are agents.  It
  is meaningless to talk about their accountability.  Even if the
  agents represented only stakeholders, Gaia says nothing about to
  whom they are accountable.  For example, if a \fsc{reviewer} may not
  send a review form, would it be responsible to the overseeing
  \fsc{pc member}?  Gaia says nothing about that.

  Zambonelli {\etal} conceptualize the multiagent system as an
  organization that ``can exploit, control or consume when it is
  working towards the achievement of the organizational goal''
  (p.~328).  Further, Gaia specifies \emph{organizational rules}
  ``that the organization should respect and enforce in its global
  behavior'' (p.~335).  Both these points seem to hint toward a
  conceptually central entity in the system.  Although Gaia never
  explicitly mentions it, perhaps all agents that adopt roles are
  accountable only to this central entity.  However, this resembles a
  conceptually centralized perspective and makes the system conception
  in Gaia similar to the one in Figure~\ref{fig:machine}.

\item \emph{Explicit Social Meaning:} Partially fulfilled.  Whereas
  Gaia supports specifying interaction among roles, it does not
  specify the meaning of the communications itself.  Gaia has the
  notions of permissions and responsibilities at the role level.  The
  intent behind these notions seems to have been to capture social
  aspects of organizations; however, Gaia falters in important
  details.  As stated above, permissions and responsibilities in Gaia
  are not interagent relationships.  For example, the \fsc{reviewer}
  has the permissions \fsf{read Papers} and \fsf{change ReviewForms}.
  But this seems to assume a computational intermediary (left
  unspecified in Gaia) with resources (the papers) where these
  activities must be performed.  (In IOSE, \fsf{change reviewForms}
  would be modeled as a communication from the \fsc{reviewer} to the
  \fsc{program chair} or the overseeing \fsc{pc member} and its body
  would contain the updated review.)  Responsibilities specify what an
  agent adopting the role ought to do but again this seems to assume
  an unspecified entity to whom the agent would be responsible.

\item \emph{Solely Social Meaning:} Violated.  As we noted above, Gaia
  does not have any true social abstractions.  

\item \emph{Separating Social from Technical:} Partially fulfilled.
  Gaia aspires to social-level modeling by modeling systems as
  organizations and having agents adopt roles in organizations.  Gaia,
  to its credit, distinguishes between open systems and closed
  systems.  It explicitly mentions that agents could represent
  different stakeholders.  However, unlike a principal in IOSE, an
  agent in Gaia is anything that has its own thread of control.
  Specifically, even in open systems, ``active'' components such as
  active databases, are modeled as agents.

\item \emph{No Principal Internals:} Violated.  Gaia makes internal
  agent behavior (e.g., \fsf{ReviewPaper}) explicit by placing
  activities in role specifications, thereby exposing an agent's
  internal decision-making. (\fsf{ReviewPaper} would not appear in an
  IOSE protocol because it does not involve communication---it is an
  internal activity.)

\end{itemize}

In summary, Gaia aspires to modeling open systems via roles and
protocols, but falters in important details.  Although it is not
explicitly machine-oriented, it betrays a conceptually centralized
mindset.  As such, it is not adequate for the modeling of open
sociotechnical systems.

\subsection{Choreographies in Service-Oriented Computing}
A choreography specifies the schemas of the messages exchanged as well
as constraints on their ordering and occurrence.  A choreography is
conceptually decentralized and involves roles that principals may
potentially adopt.  Interestingly, choreographies fit in the paradigm
of Figure~\ref{fig:old}.  Principals adopting roles in a choreography
would potentially maintain their own ledgers and interact via a
digital messaging system that carries their messages to each other.
However, there is no computational support for the social state:
choreographies do not specify the social meanings of messages and as a
result do not capture the social expectations interacting principals
may have of each other.

Choreography description languages, e.g., WS-CDL \cite{WS-CDL:2005},
support specifying the internal activities of principals in the
choreography itself.  For instance, the WS-CDL notion of
\emph{workunits} may be used to specify conditional actions by a
principal.  Likewise, Mendling and Hafner
\shortcite{mendling:ws-cdl:2008} use workunits to specify the internal
compliance checks that a tax adviser would make in handling a client's
annual statement (p.~532), thereby violating encapsulation.

\begin{itemize}
\item \emph{Accountability Modularity:} Partially fulfilled.
  Choreographies support compliance at the technical level: principals
  must send messages in the prescribed order, otherwise they are not
  compliant with the choreography.  However, lacking representation of
  social meaning and social state, a violation of a choreography does
  not necessarily amount to a violation at the social level.

\item \emph{Explicit Social Meaning:} Violated.  Lack a representation
  of social meaning.

\item \emph{Solely Social Meaning:} Violated.  Follows from the lack
  of a representation of social meaning.

\item \emph{Separating Social from Technical:} Fulfilled.
  Choreographies specify interactions with reference to roles that
  principals may adopt.

\item \emph{No Principal Internals:} Partially fulfilled.  In
  principle, choreographies seek to specify interactions; however, in
  practice, they also specify the internal activities of principals.

\end{itemize}

Choreographies seek to model interactions; however, they do so in
terms of low-level control and data flow abstractions and often
specify aspects of principals' internal behavior.  Therefore, they
also do not fare well with respect to the IOSE principles.

\subsection{Summary of Evaluation}
\label{subsec:summary-eval}
Table~\ref{table:principles-evaluation} contrasts traditional
approaches with commitment protocols, explained in
Section~\ref{sec:principles-IOSE} as an exemplar of IOSE.  i*, Tropos,
and KAOS are all machine-oriented since their primary specification
artifact is that of a machine (the system actor in i* and Tropos and a
collection of agents in KAOS) that would meet stakeholder
requirements.  Nonetheless, i* and Tropos perform better than KAOS in
our evaluation because unlike KAOS, they model stakeholders and their
relations explicitly.


\begin{table}[htb!]
  \centering \tbl{Evaluation: {\F}, {\Q}, and {\V} stand for
    fulfilled, partially fulfilled, and violated,
    respectively.\label{table:principles-evaluation}}{%
\begin{tabular}{lccccc}  \toprule
\fbf{Principle}   & \fbf{i* \& Tropos}  & \fbf{KAOS} & \fbf{Gaia} & \fbf{Choreographies} & \fbf{Commitments}\\ \midrule
Accountability modularity          & \V & \V         & \V     &\Q     & \F \\
Explicit social meaning            & \Q & \V         & \Q     &\V     & \F \\
Solely social meaning              & \Q & \V         & \V     &\V     & \F \\
Separating social from technical   & \Q & \V         & \Q     &\F     & \F \\
No principal internals             & \V & \V         & \V     &\Q     & \F \\
\bottomrule
\end{tabular}
}
\end{table}

\section{Relevant Literature}
\label{sec:literature}


In the foregoing, we have extensively discussed the literature most
pertinent to our claims.  Here, we discuss other relevant literature.

\mypara{Commitments.}  Commitments are recognized in the
literature as a key social abstraction.  Singh
\shortcite{singh:rethinking:1998} introduced social commitments as a
way of formalizing agent communication that was suited to open
systems.  Fornara and Colombetti
\shortcite{fornara:operational-commiments:2002} gave an operational
semantics for commitments.  Yolum and Singh
\shortcite{yolum:flexible-protocols:2002} specified commitment
protocols in the event calculus.  Newer work has tended more toward SE
themes: methodologies for composing commitment protocols
\cite{desai:amoeba:2010}, patterns
\cite{scd:CSOA:2009,chopra:patterns:2011}, the relationship of
commitments with the notion of goals in Tropos
\cite{chopra:goals-com-caise:2010}, and monitoring requirements via
commitments \cite{robinson:commitments:2009}.  Young and Ant\'{o}n
\shortcite{young:commitments:2010} apply commitments to identify
software requirements from regulatory policies and Paja {\etal}
\shortcite{paja:sts-security:2012} to extract security requirements
for organizations.  Baldoni {\etal}
\shortcite{baldoni:regulative:2012} present commitment protocols that
support temporal properties.

The value we add to this body of work is making explicit connections
between the ontology and principles implicit in commitment-based
approaches with those implicit in traditional SE.  IOSE, as a new
paradigm for the sociotechnical systems, provides a natural home for
commitments.

\mypara{Agent-Oriented Software Engineering}  Despite the apparent
similarity between the notion of agents and principals and talk of
interaction in many AOSE methodologies, they (as exemplified by Gaia
and Tropos) violate key IOSE principles.  Some AOSE approaches are
logically centralized, e.g., geared toward efficient problem-solving.
Gaia and other AOSE methodologies acknowledge the distinction between
open and closed systems and emphasize interactions.  They correctly
classify problem solving systems that distribute a problem across
agents as closed.

However, AOSE methodologies betray, if not centralized mindsets, at
least considerable conceptual difficulties.  V\'{a}squez-Salceda
{\etal} \shortcite{vasquez-salceda:organizations:2005} model the
objectives of social systems as goals and the systems as controllers
(p.~338): ``Facilitation roles are usually provided by agents
controlled by the society, and follow a trivial contract.''  But in a
sociotechnical system, there is no unitary multiagent system or
society that controls resources.  And an organization has no goals
that overarch the goals of other participants---that is, an
organization would be one of many participants in a sociotechnical
system.  Van Riemsdijk {\etal} \shortcite{vanRiemsdijk:orgs:2009} too
are motivated by the modeling of open systems but conceptualize a
multiagent system as having collective goals.

Wagner's \shortcite{wagner:aor:2003} Agent-Object-Relationship (AOR)
ontology proposes two kinds of models to support the design of a
multiagent system: \emph{external} (observer's perspective,
including interaction models) and \emph{internal} (agent's
perspective).  Newer work by Guizzardi {\etal}
\shortcite{guizzardi:ontology:2010} is based on the AOR ontology.
Wagner's internal-external distinction aligns well with IOSE.
Unfortunately, AOR subverts the internal-external distinction by
including in external models aspects of the internal organization of
agents, such as their beliefs (in \emph{agent diagrams}) as well as
reactive rules that specify how an agent would respond to events.
Singh \shortcite{singh:distinction:1991} introduced social commitment
as a concept distinct from the mentalist concept of \emph{internal
  commitment} \cite{cohen:intentions:1990}.  AOR, Tropos, and i*
employ internal commitments.

Serrano and Leite \shortcite{serrano:istar:2011} map i* concepts to
agent programming concepts such as beliefs, desires, and intentions.
They map dependencies between two agents to the joint desires of those
agents.  Using their methodology, one specifies the internals of one
or more agents.  Thus even though Serrano and Leite don't involve an
explicit system actor, the approach is conceptually centralized.

How can we reconcile the facts that many AOSE methodologies are
conceptually centralized yet model interactions?  AOSE methodologies
yield physically distributed systems of agents that pursue centrally
assigned goals.  Interactions in AOSE are focused on the technical
means to achieve such goals whereas interactions in IOSE have social
standing.

\mypara{Information Systems Modeling} McCarthy
\shortcite{mccarthy:rea:1982} proposed REA as a generalized model for
accounting that is based on \emph{resources}, \emph{events}, and
\emph{agents} and relations among them.  REA is finding applications
in services and information systems modeling, e.g.,
\cite{weigand:rea:2011}.  REA seeks to capture the fundamental
economic model underlying accounting.  Our endeavor is analogous: to
develop a general model of sociotechnical systems, based on social
abstractions.  The key distinction between IOSE and REA is that REA
models ultimately represent an organization's internal perspective,
not its interactions with others.  This is evident from REA's
inside-outside distinction.  One of REA's \emph{accountability}
relations identifies the responsible agent for an event \emph{inside}
an organization; another is between an outside agent and an event.
IOSE commitments are more general: each commitment identifies the
accountable party and the party to whom it is accountable for what.

The second distinction stems from the \emph{duality} of events and the
corresponding \emph{reciprocity} of commitments---reflecting the roots
of REA in accounting.  Duality in REA relates two events that together
characterize the give-and-take of an economic exchange (e.g.,
\fsf{pay} and \fsf{deliver}).  These events \emph{execute} (i.e.,
fulfill) the corresponding reciprocal commitments.  IOSE does not give
any special place to either duality or reciprocity.  No IOSE
commitment is necessarily reciprocal: If $x$ is committed to $y$,
there is no requirement that $y$ be committed to $x$ (though it is
allowed).

Gordijn and Akkermans \shortcite{gordijn:ebusiness-models:2001}
propose the $e^3$-value ontology and associated methodology in order
to specify and analyze so-called business models to evaluate their
profitability for all stakeholders involved.  A business model
represents stakeholders as actors (e.g., customer).  Other important
concepts are \emph{value objects} (e.g, online articles), and
\emph{value exchanges} among actors (e.g., with an article provider
for articles and their payment).  Commitment protocols could be used
to capture the notion of value exchange among actors; further
commitment protocols could be analyzed for feasibility
\cite{desai:contracts:2008}.  However, unlike commitment protocols and
in violation of IOSE principles, $e^3$-value models also depict the
internal flow of values in Petri Net-like notation.

\mypara{Human-Computer Interaction (HCI)} Following developments in
HCI, the term ``social'' is often associated in RE with the interplay
between social factors and the design of technical artifacts (as in
the \emph{design of work}, e.g., see \cite{flores:coordinator:1988}),
especially as concerns user interface and experience (e.g., see
\cite{sutcliffe:user-interaction-2011}).  The IOSE treatment of
``social'' is more general and includes not just people but entities
such as enterprises and organizations that may be conceptualized as
social actors---which enables IOSE to tackle sociotechnical systems as
we define them.

\mypara{Compliance}  Compliance with service-level agreements,
regulations, and business contracts is gaining interest in software
engineering.  Organizations are increasingly concerned with the
question of compliance with regulatory frameworks such as HIPAA
\cite{young:commitments:2010} and Sarbanes-Oxley.  The challenges here
are threefold: one, how to model regulations; two, how to determine
the compliance of a principal's behavior with regulations; and, three,
how to design a principal's information systems such that it is
(likely to be) compliant with the regulations.  Siena {\etal}
\shortcite{siena:nomos:2010} develop a Tropos-inspired modeling
language to capture and reason about such regulations.  IOSE can
potentially support a more flexible notion of compliance geared toward
open systems.

Although high-level concepts such as obligations, permissions, and so
on are often employed in software engineering, they are used to
capture behavioral constraints abstractly but are not grounded in
communication as commitments are.  A more detailed discussion is out
of the scope of this paper but interested readers are referred to
\cite{Castelfranchi-Festschrift-12} for an in-depth discussion.


\section{Discussion and Future Work}
\label{sec:discussion}
Our work began from the recognition that open sociotechnical systems
are conceptually decentralized and their principals are autonomous
social entities.  Specifying a sociotechnical system amounts to
specifying a protocol, i.e., a specification of the interactions among
its potential participants.  We showed that existing approaches take a
fundamentally centralized view of sociotechnical systems design: they
are geared toward the specification of a machine, not a protocol.

We presented IOSE as a new approach for the engineering of
sociotechnical systems.  IOSE is driven by key technical demands:
autonomy, accountability, and loose coupling.  To illustrate IOSE, we
introduced protocols in which the meanings of the messages are
specified in terms of commitments as a natural way to address these
technical demands.  We formulated the IOSE principles inspired by, but
distinct from, the traditional SE principles.  We undertook a detailed
analysis of some leading relevant SE approaches to show which IOSE
principles they violate.  Traditional SE falls short not because the
abstractions it employs are low level, but because they deemphasize
interaction.  To support this claim, we consider well-known
requirements modeling approaches, which emphasize high-level
abstractions such as goals and dependencies
\cite{mylopoulos:goals:1999} but do not accord first-class status to
interactions.

Even in the simple appointment scheduling setting, the benefits of
IOSE are obvious: what was previously seen as the problem of
specifying a meeting scheduler, under IOSE turns into two independent
problems of (1) specifying the meeting scheduling protocol, and (2)
specifying principals to participate in the protocol.  This reflects
Parnas' conception of modularity as the \emph{division of labor}.

Our evaluation of the traditional approaches is not a blanket
criticism but rather a critical examination that highlights the
differences in their assumptions from the principles of IOSE.
Further, our evaluation was limited to the system models or
specifications that existing approaches advocate.  We did not, for
instance, dwell upon the modeling of functional versus nonfunctional
requirements or the elicitation of requirements from stakeholders,
areas in which current sociotechnical systems approaches (including
i*, Tropos, and KAOS) have made substantial progress.

Although we carefully differentiate IOSE from traditional SE, the two
complement each other.  IOSE deals with systems (viewed as protocols)
whereas traditional RE deals with components or principals' behaviors
(viewed as machines).  The former is typified by commitments and other
social constructs; the latter by goals and other mental constructs.
Recent works \cite{chopra:goals-com-aamas:2010,ProMAS-11} have begun
to relate these views from the standpoint of building agents who can
function effectively in sociotechnical systems characterized by
commitments.  We have recently
\cite{chopra:goals-com-caise:2010,telang:tropos:2009}, sought to apply
Tropos-like models towards the specification of sociotechnical
systems.  These works replace intentional dependencies with
commitments and do not consider a system actor.  Further, each actor
is understood as a role, i.e., inert and lacking any goals.  Goals are
applied in the modeling the principals who may potentially adopt roles
in the sociotechnical system.

We defer the task of systematically laying out the methodological
elements of IOSE, as they overlay the main concepts and principles we
introduced above.  We expect to go beyond methodologies for specifying
and composing protocols, e.g., \cite{desai:amoeba:2010}, by
incorporating the key RE considerations of stakeholders and
requirements.  The nature of requirements for protocols appears quite
different from requirements for machines.  For example, a requirement
such as ``a meeting is scheduled within three hours of the request''
makes sense for a machine but not a protocol---though it could be
negotiated between the principals playing the \fsc{mi} and \fsc{mp}
roles.  A requirement that ``a principal can observe whether another
principal is complying with some commitment'' makes sense for a
protocol but less so for a machine.  The bases from which to engineer
protocols is a timely and pertinent direction of research.  We imagine
that existing RE methodologies would provide a useful starting point
for investigations into a methodology for IOSE.

\section*{Acknowledgments}
Amit Chopra was supported by a Marie Curie Trentino Cofund grant and
the ERC Advanced Grant 267856.  We thank Fabiano Dalpiaz, Paolo
Giorgini, Michael Huhns, Michael Jackson, John Mylopoulos, and Erik
Simmons for helpful discussions.


\end{document}